\documentclass[amsmath,amssymb,floatfix,aps,prd,11pt,tightenlines,preprintnumbers,nofootinbib,prd]{revtex4-2}   

\input{header}
\usepackage{comment}
\usepackage{lineno}
\usepackage{listings}
\usepackage{bm}
\usepackage{orcidlink}
   
\newcommand{\openingAngle}{\psi}
\newcommand{\openingAngleCapital}{\Psi}
\newcommand{\zee}{\xi}

\makeatletter
\def\l@subsection#1#2{}
\def\l@subsubsection#1#2{}
\makeatother

\begin{document}

%+++++++++++++++++++++++++++++++++++++++
%TitlePage
%+++++++++++++++++++++++++++++++++++++++
\preprint{UCI-TR-2025-04}

\title{Spin Correlations in Dark Photon Searches \vspace*{0.15in}}

\author{Jonathan L.~Feng\,\orcidlink{0000-0002-7713-2138}}
\email{jlf@uci.edu}
\affiliation{Department of Physics and Astronomy, University of California, Irvine, CA 92697, USA \vspace*{0.05in} }

\author{Mi\v sa Toman\,\orcidlink{0009-0003-2647-1281}}
\email{mtoman@uci.edu}
\affiliation{Department of Physics and Astronomy, University of California, Irvine, CA 92697, USA \vspace*{0.05in} }

\author{Eli Welch\,\orcidlink{0000-0001-6336-2912}}
\email{welchem@uci.edu}
\affiliation{Department of Physics and Astronomy, University of California, Irvine, CA 92697, USA \vspace*{0.05in} }

\begin{abstract}
We investigate the spin correlations between production and decay in the process where dark photons $A'$ are produced in pseudoscalar meson decays, for example, $\pi^0, \eta \to \gamma A'$, and then decay to fermion pairs, $A' \to f \bar{f}$.  This process is the focus of many experimental searches, but spin correlations are typically ignored. We derive analytic results that allow us to quickly scan parameter space and quantify the error made in neglecting spin correlations.  In particular, we define a discrepancy parameter and find that this parameter is always less than $\frac{1}{6\sqrt{3}} \approx 9.6\%$, which provides a rough measure of the size of spin correlation effects.  However, these effects may be enhanced or suppressed by cuts in realistic experimental analyses, and so we also consider two representative examples, including current FASER analyses and possible future searches at SHiP.  We find that the effects of spin correlations are negligible for existing FASER analyses, but can significantly reduce event rates at upcoming SHiP searches. 
\end{abstract}

\maketitle

\vspace*{-0.2in}
\tableofcontents

%****************************************
\section{Introduction}
\label{sec:intro}
%****************************************

In the search for physics beyond the standard model (BSM), long-lived particles have become important targets.  Among these, the dark photon $A'$~\cite{Holdom:1985ag} is especially prominent.  Dark photons appear when there is a dark sector with an Abelian gauge symmetry, and the dark gauge boson mixes with the standard model (SM) photon through a renormalizable ``portal'' interaction. For the unprobed parameter space with dark photon masses $m_{A'} \sim 10~\mev - \gev$ and couplings $\varepsilon \sim 10^{-6} - 10^{-3}$, dark photons are naturally embedded in models with thermal relics with the right abundance to be a significant fraction (or all) of dark matter~\cite{Boehm:2003hm, Pospelov:2007mp, Feng:2008ya}.  Such masses and couplings therefore merit extra attention.  Recent searches at a variety of experiments, including NA64~\cite{NA64:2019auh}, the Heavy Photon Search~\cite{Adrian:2022nkt}, FASER~\cite{FASER:2023tle}, and NA62~\cite{NA62:2023nhs}, have probed this parameter region, and searches with rapidly-increasing sensitivity are expected in the near future.

For masses in the 10 MeV to GeV range, dark photons are often dominantly produced in pseudoscalar meson decays, such as $\pi^0 , \eta \to \gamma A'$, and then decay to fermion pairs $A' \to f \bar{f}$, where $f = e, \mu, \ldots$.  Because the dark photon is long-lived, traveling a macroscopic distance between production and decay, the narrow-width approximation is valid, and it is tempting to treat the production and decay processes independently.  For example, in FORESEE~\cite{Kling:2021fwx}, a leading event generator used to simulate signal events at forward experiments at the Large Hadron Collider (LHC), dark photons are produced in one module, propagated, and then decayed in another module, neglecting the spin correlations between production and decay.  Of course, this is formally incorrect, as the dark photon is not a scalar, and spin correlations are present, although, to our knowledge, they have been included for the process we consider in only one analysis~\cite{Kahn:2014sra}.  There are notable examples where spin correlations between production and decay have been shown to be important, even when the narrow-width approximation is valid.  For example, spin correlations are critically important in processes involving heavy neutrinos, and they can even shed light on their Dirac or Majorana nature~\cite{Han:2012vk}.

In this study, we derive analytical results for the dark photon processes $\pi^0, \eta \to \gamma A' \to \gamma f \bar{f}$, including the spin corrections between production and decay.  For concreteness, we will often speak of the representative process $\pi^0 \to \gamma A' \to \gamma e^+ e^-$, but our analysis is valid for arbitrary meson, dark photon, and fermion masses.  The only approximation we make in our calculation is the narrow-width approximation, which, of course, is highly accurate, given that the dark photon is a long-lived particle.  Our analytic results allow us to quickly scan parameter space and determine, for example, what dark photon and fermion masses are expected to amplify the effects of spin correlations.  We also show how the effects of spin correlations may be understood as resulting from the anisotropic emission of electron-positron pairs in the dark photon frame.  

As an application of our analytic results, we calculate the emitted electron/positron energy spectra with and without spin correlations in both the meson center-of-mass (COM) frame, which we will refer to as the meson frame or pion frame, and in the lab frame. To quantify the difference between the spectra with and without spin correlations, we define the discrepancy parameter
\begin{equation}
\Delta  = \frac{1}{2 N_e} \int_{E_{\text{min}}}^{E_{\text{max}}}
\left| \left(\frac{dN_e}{dE_e}\right)_{\text{with}} - \left( \frac{dN_e}{dE_e} \right)_{\text{without}} \right| d E_e \ ,
\label{eq:Delta}
\end{equation}
which is 0 when the spectra are identical, and 1 when the spectra have no overlap at all.  For the dark photon process we consider, we find that $\Delta \le \frac{1}{6\sqrt{3}} \approx 9.6\%$ over the entire range of dark photon parameters, which provides a rough estimate of the possible size of spin correlation effects.  

The parameter $\Delta$ is Lorentz invariant.  However, once (frame-dependent) energy and angular cuts and other experimental details are included, the effect of spin correlations can either be enhanced or suppressed.  We therefore also analyze the effects of spin correlations in two representative cases, including a current analysis deriving leading bounds on dark photons from FASER at the LHC~\cite{FASER:2022hcn}, and a potential future dark photon search at the fixed target experiment SHiP at the Super Proton Synchrotron (SPS)~\cite{Albanese:2878604}.  We find that spin correlations are negligible for existing FASER analyses, but can significantly reduce the number of signal events passing cuts in upcoming SHiP analyses.  Given expected improvements in sensitivity, spin correlations cannot be neglected in future analyses.

The paper is structured as follows. In \cref{sec:technical-intro} we discuss dark photon production and decay in general, and identify the source of spin correlations based on a simple argument using angular momentum conservation.  In \cref{sec:feynman-amplitudes} we establish our notation and conventions, and we calculate the squared Feynman amplitudes, $| {\cal M}|^2$, for the process $\pi^0 \to \gamma A' \to \gamma e^+ e^-$ with and without spin correlations.  We then determine the fermion energy distributions in the meson and lab frames with and without spin correlations in \cref{sec:differential-decay-width-in-the-meson-com}.  In particular, we derive a remarkably simple analytical expression for the discrepancy parameter $\Delta$ of \cref{eq:Delta}.  In \cref{sec:angulardistributions}, we discuss the fermion angular distributions in the dark photon, meson, and lab frames.  In \cref{sec:the-impact-of-spin-correlations} we quantify the importance of spin correlations in the two representative experimental analyses at FASER and SHiP, as discussed above.  We summarize our conclusions in \cref{sec:conclusions}. 

In this work, at times it will be useful to phrase the results in terms of probability distributions, and also to use probability density calculus to boost these distributions from one frame to another.  In \cref{appendix:pdf-calculus-intro}, we summarize the essentials of probability density calculus used in this paper.  We then apply it to boost probability densities from one frame to another in two examples in \cref{appendix:pdf-calculus-1to2-decay,appendix:pdf-calculus-three-body}, and show that the results without spin correlations can be derived using only four-momentum conservation and classical physics. Last, in \cref{appendix:why-is-there-a-plateau?}, we explain the physical meaning behind the distinctive features of the fermion energy spectrum in the lab frame.

%****************************************
\section{Dark Photon Production and Decay}
\label{sec:technical-intro}
%****************************************

Dark photons are gauge bosons arising from a broken $U(1)_D$ symmetry in the dark sector. The only gauge-invariant and renormalizable interaction term between SM particles and the dark photon is the kinetic mixing term $B_{\mu\nu}F'^{\mu\nu}$, where $B_{\mu\nu} = \del_\mu B_\nu - \del_\nu B_\mu$ is the field strength tensor of the $U(1)_Y$ hypercharge gauge boson and $B_\mu = -\sin\theta_W Z_\mu + \cos\theta_W A_\mu$ is a mixture of the $Z$ boson and the photon $A$, and $F'^{\mu\nu}$ is the U(1)$_D$ gauge field strength tensor. In the limit $m_D\ll m_Z$, where $m_D$ is the mass of the U(1)$_D$ gauge boson, the mixing with the $Z$ boson can be neglected~\cite{Feldman:2007wj,Davoudiasl:2012ag,Feng:2016ijc}, and the dark gauge boson effectively mixes only with the SM photon through the term $F_{\mu\nu} F'^{\mu\nu}$.  This mixing between the two gauge bosons can be removed by a change of basis. The end result is the Lagrangian 
\begin{align}
\mathcal{L} & = \overline{f}\left(i\slashed{\del} - m_f\right) f + \varepsilon e Q_f \Bar{f}\slashed{A'}f + \frac{1}{2} m^2 A'_\mu A'^\mu - \frac{1}{4} F^{\mu\nu} F_{\mu\nu} - \frac{1}{4} F'^{\mu\nu} F'_{\mu\nu} \, ,
\label{mixing-term:fermions-darkphoton}
\end{align}
where $Q_f$ is the electric charge of SM fermion $f$, which extends the SM by adding a dark photon $A'$ with mass $m$, which couples to SM fermions just like the photon, but with an extra factor of $\varepsilon$. 

To couple pseudoscalar mesons, like the neutral pion $\pi^0$, to photons and dark photons, we recall that pseudoscalars couple to a pair of photons through the axial anomaly term~\cite{Weinberg:1996kr}
    \begin{align}
        \mathcal{L}_{\pi\gamma\gamma} & = -g \pi^0\varepsilon^{\mu\nu\rho\sigma}F_{\mu\nu}F_{\rho\sigma}\,,\quad g = \frac{N_c e^2}{48\pi^2 f_\pi}\,,\quad f_\pi\approx 184~\text{MeV}\, .
    \end{align}
The Feynman rule associated with this interaction vertex is
    \begin{align}
    8ig\varepsilon^{\mu\nu\rho\sigma}p_\rho q_\sigma\,,
    \end{align}
and the pion decay width is
    \begin{align}
        \Gamma\left(\pi^0\to2\gamma\right) = \frac{g^2}{\pi}M_\pi^3\,.
    \end{align}

Dark photons interact with pions in the same way, but suppressed by the kinetic mixing parameter $\varepsilon$,
    \begin{align}\label{mixing-term:photon-darkphoton}
        \mathcal{L}_{\pi\gamma A'} & = -2\varepsilon g\pi^0\varepsilon^{\mu\nu\rho\sigma}F_{\mu\nu}F'_{\rho\sigma}\, ,
    \end{align}
where the extra factor of 2 appears because we no longer have two identical particles in the final state. The Feynman rule associated with this interaction vertex is 
    \begin{align}\label{vertex}
    8i\varepsilon g\varepsilon^{\mu\nu\rho\sigma}p_\rho q_\sigma\,, 
    \end{align}
and the resulting decay width is
    \begin{align}
        \Gamma(\pi^0\to \gamma A') & = \frac{2\varepsilon^2 g^2}{\pi} M_\pi^3\left(1 - \frac{m^2}{M_\pi^2}\right)^3\,.
    \label{eq:mesondecaywidth}
    \end{align}
The branching ratio for $\pi^0\to\gamma A'$ is, then~\cite{Batell:2009di},
\begin{align}
    B\left(\pi^0\to \gamma A' \right) & = 2\varepsilon^2\left(1 - \frac{m^2}{M_\pi^2}\right)^3B\left(\pi^0\to\gamma\gamma\right)\,,
\end{align}
with analogous formulas for dark photons produced in the decays of $\eta$ and other pseudoscalar mesons.

Once produced, the dark photon decays to fermion--anti-fermion pairs with decay width 
\begin{align}\label{eq:darkphotondecaywidth}
        \Gamma\left(A'\to f\bar f\right) & = \frac{\varepsilon^2e^2}{12\pi}m \sqrt{1-\frac{4m_f^2}{m^2}}\left(1 + \frac{2m_f^2}{m^2}\right)\,.
    \end{align}
The decay width is highly suppressed for couplings in the range $\varepsilon \sim 10^{-6} - 10^{-3}$.  The momentum dependence of the $\pi^0\gamma A'$ interaction vertex, shown in \cref{vertex}, may seem to invalidate the narrow-width approximation, as discussed in Ref.~\cite{Berdine:2007uv}. However, the extremely small value of the decay width justifies the use of the narrow-width approximation.

Before deriving analytic results for dark photon events with and without spin correlations, with the results presented above, we can already identify the source of the spin correlations:~the decay $\pi^0\to\gamma A'$ to a longitudinally-polarized $A'$ is forbidden by angular momentum conservation. This can be seen from the Feynman amplitude for the decay in the pion frame. For photon momentum $p$ and polarization $r$ and dark photon momentum $q$ and polarization $l$, this amplitude is
\begin{align}
    \mathcal{M}\left(\pi^0\to\gamma A'\right) & \propto \varepsilon^{\mu\nu\rho\sigma}p_\rho q_\sigma\varepsilon_\mu^{(r)}(p)\varepsilon^{(l)}_\nu(q) = \varepsilon^{\mu\nu\rho\sigma}(k_\rho - q_\rho) q_\sigma\varepsilon_\mu^{(r)}(p)\varepsilon^{(l)}_\nu(q)\nonumber\\
    & = \varepsilon^{\mu\nu\rho\sigma}k_\rho q_\sigma\varepsilon_\mu^{(r)}(p)\varepsilon^{(l)}_\nu(q) = \varepsilon^{\mu\nu0\sigma}M_\pi q_\sigma\varepsilon_\mu^{(r)}(p)\varepsilon^{(l)}_\nu(q)\nonumber\\
    & = M_\pi\boldsymbol{\varepsilon}^{(r)}(p)\cdot\left(\boldsymbol{\varepsilon}^{(l)}\times\bm{q}\right) = 0 \ .
\end{align}
The first line uses 4-momentum conservation; the second line follows from working in the pion frame, where $k^\rho = (M_\pi, 0, 0, 0)$; and the vanishing of the amplitude in the third line follows from the fact that the dark photon momentum $\bm{q}$ is parallel to its longitudinal polarization vector. If spin correlations are neglected, the dark photon is decayed assuming it is unpolarized, which neglects the fact that the dark photon is produced transversely polarized in pion decay.  This fact is, of course, accounted for in a full treatment of the process with spin correlations included.

%****************************************  
\section{Squared Feynman Amplitudes}
\label{sec:feynman-amplitudes}
%****************************************

We now calculate the squared Feynman amplitudes, $| {\cal M}|^2$, for the process $\pi^0 \to \gamma A' \to \gamma e^+ e^-$, with and without spin correlations in the narrow-width approximation.  We will use the conventions of Peskin and Schroeder~\cite{Peskin:1995ev}. As noted in \cref{sec:intro}, we will use pions as our representative pseudoscalar mesons and electrons as our representative fermions, but the analysis is valid for arbitrary pseudoscalar meson, dark photon, and fermion masses.

\begin{figure}[tbp]
%\feynmandiagram [layered layout, horizontal=a to b] {
%        a [particle=\(\pi^0\,\eta\)] -- [scalar, momentum'={\(k\)}] b -- [boson, momentum={\(p\)}] b1 [particle=\(\gamma\left(r\right)\)],
%        b -- [boson, momentum'={\(q\)}, edge label=\(A'\)] c,
%        c -- [anti fermion, momentum={\(k_2\)}] f2 [particle=\(\overline{f}\left(s_2\right)\)],
%        c -- [fermion, momentum'={\(k_1\)}] f3 [particle=\(f\left(s_1\right)\)]
%        }; 
\includegraphics[width=0.5\textwidth]{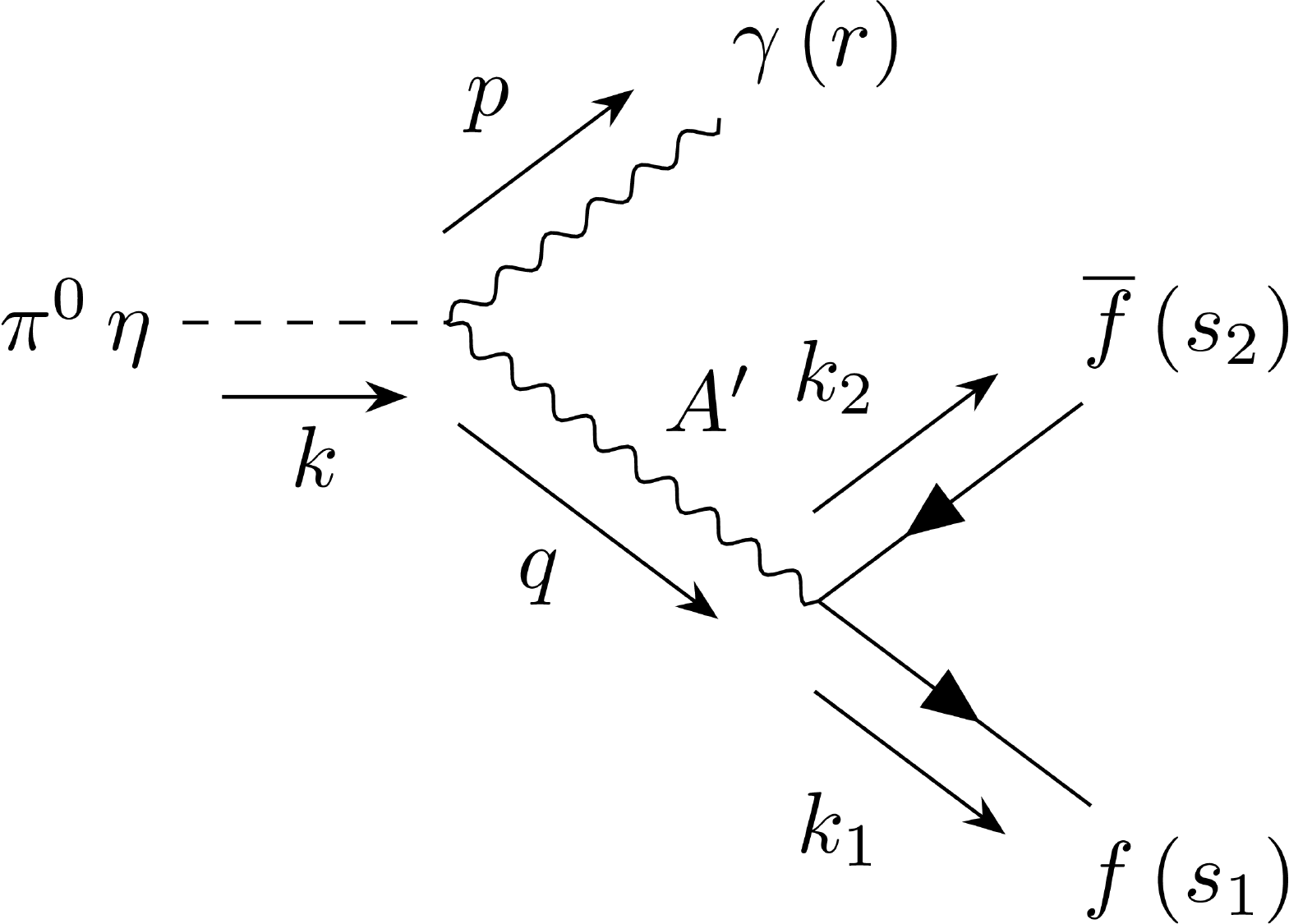}
\caption{Feynman diagram for the production and decay of dark photons through $\pi^0, \eta \to \gamma A' \to \gamma \bar{f} f$. The masses of the decaying meson, dark photon, and fermions are denoted $M$, $m$, and $m_f$, respectively.  The four-momenta of the particles are $k = \left(E_{\bm{k}}, \bm{k}\right)$, $p = \left(E, \bm{p}\right)$, $q = \left(E', \bm{q}\right)$, $k_1 = \left(E_1, \bm{k}_1\right)$, and $k_2 = \left(E_2, \bm{k}_2\right)$. The polarizations of the final-state particles are $r$, $s_1$, and $s_2$, as labeled. }
\label{fig:pion-decay}
\end{figure}

Our notation is summarized in the Feynman diagram shown in \cref{fig:pion-decay}.  When the narrow-width approximation is valid, the pion-frame energies of the photon and dark photon are 
\begin{align}
    E = \frac{M^2 - m^2}{2M} \quad \text{and} \quad 
    E' = E_1 + E_2 = \frac{M^2 + m^2}{2M}\,,
    \label{eq:energies}
\end{align}
respectively. 

%****************************************
\subsection{Squared Feynman Amplitude with Spin Correlations}\label{subsec:Feynman amplitude with spin correlation}
%****************************************

The Feynman amplitude for the process $\pi^0 \to \gamma A' \to \gamma e^+ e^-$ shown in \cref{fig:pion-decay} is 
\begin{align}\label{eqn:main-feynman-amplitude}
    i\mathcal{M} & = i\varepsilon g M^\nu_r(p,q)\cdot\frac{-iP_{\nu\alpha}}{q^2 - m^2 - im \, \Gamma(A')}\cdot\left( -i\varepsilon e N^\alpha_{s_1, s_2}(k_1, k_2)\right)\,,
\end{align}
where
\begin{align}
    M^\nu_r(p,q) & = 8\varepsilon^{\mu\nu\rho\sigma}p_\rho q_\sigma\varepsilon_\mu^{(r)}(p)\,,\quad P_{\nu\alpha} = g_{\nu\alpha} - \frac{q_\nu q_\alpha}{m^2}\,,\quad N^\alpha_{s_1, s_2}(k_1, k_2) = \bar{u}_{s_1}(k_1)\gamma^\alpha v_{s_2}(k_2)\,.
\end{align}

We assume that the final-state particle polarizations are not measured, and so, averaging over initial-state polarizations (of which there is only one in this case) and summing over final-state polarizations, we find
\begin{align}
\text{Av} \! \! \sum_{r, s_1,s_2} \!
\left| \mathcal{M} \right|^2
= g^2 e^2\varepsilon^4\underbrace{\sum_r M^{*\nu}_r M^{\nu'}_r}_{M^{\nu\nu'}} P_{\nu\alpha}P_{\nu'\alpha'}\overbrace{\sum_{s_1,s_2}N^{*\alpha}_{s_1,s_2}N^{\alpha'}_{s_1,s_2}}^{N^{\alpha\alpha'}}\frac{1}{\left|q^2 - m^2 - im \, \Gamma(A')\right|^2}\,,
\end{align}
where
\begin{align}
M^{\nu\nu'} & = \sum_r M^{*\nu}_r M^{\nu'}_r = 64\varepsilon^{\mu\nu\rho\sigma}\varepsilon^{\mu'\nu'\rho'\sigma'}p_\rho q_\sigma p_{\rho'}q_{\sigma'}\sum_r\varepsilon^{*(r)}_{\mu}(p)\varepsilon^{(r)}_{\mu'}(p) \nonumber \\
& = -64\varepsilon^{\mu\nu\rho\sigma}\varepsilon\indices{_\mu^{\nu'}^{\rho'}^{\sigma'}}p_\rho q_\sigma p_{\rho'}q_{\sigma'} \\ N^{\alpha\alpha'} & = 4\left[ k_2^\alpha k_1^{\alpha'} + k_1^{\alpha} k_2^{\alpha'} - \left(k_2\cdot k_1 + m_f^2\right)g^{\alpha\alpha'}\right] .
\end{align}
Given the anti-symmetric 4-index tensor in $M^{\nu\nu'}$, the propagator factors act simply as raising and lowering operators: $M^{\nu\nu'}P_{\nu\alpha} = M^{\nu\nu'}g_{\nu\alpha} = M\indices{_\alpha^{\nu'}}$, and $M^{\nu\nu'}P_{\nu\alpha}P_{\nu'\alpha'} = M_{\alpha\alpha'}$. We find, then, that
\begin{align}
M^{\nu\nu'} P_{\nu \alpha} P_{\nu' \alpha'} N^{\alpha\alpha'} &=
M_{\alpha\alpha'}N^{\alpha\alpha'} \nonumber \\
& = 256\left[g_{\alpha\alpha'} p^2q^2 + q_\alpha p_{\alpha'}p\cdot q + p_{\alpha}q_{\alpha'}p\cdot q - g_{\alpha\alpha'}\left(p\cdot q\right)^2 - q_\alpha q_{\alpha'}p^2 - p_{\alpha}p_{\alpha'}q^2\right] \nonumber \\
& \qquad \times \left[ k_{2}^{\alpha} k_{1}^{\alpha'} + k_{1}^{\alpha} k_{2}^{\alpha'} - \left(k_2\cdot k_1 + m_f^2\right) g^{\alpha\alpha'}\right] \nonumber \\
&= 512\left[ A - q^2\Dot{k}{k_2}\Dot{k}{k_1}\right] \, ,
\end{align}
where $A = \Dot{p}{q}^2 ( \frac{1}{2}q^2 + m_f^2 ) + \frac{1}{4} q^4 M^2$ becomes momentum-independent once the dark photon is put on shell.  With this, we derive
\begin{align}\label{eqn:main-feynman-amplitude-squared-with-spin-correlations}
\text{Av} \! \! \sum_{r, s_1,s_2} \! \left| \mathcal{M} \right|^2
&= 512 \, g^2 \, e^2 \, \varepsilon^4 \, \frac{A - q^2\Dot{k}{k_2}\Dot{k}{k_1}}{\left|q^2 - m^2 - im \, \Gamma(A') \right|^2}\,.
\end{align}

Applying the narrow-width approximation to the dark photon propagator, 
\begin{align}
    \frac{1}{\left|q^2 - m^2 - im \, \Gamma(A')\right|^2} & \to \frac{\pi}{m \, \Gamma(A')}\delta\left(q^2 - m^2\right)\,,
\end{align}
which puts the dark photon on shell, setting $q^2 = m^2$, we find that the averaged squared amplitude is
\begin{align}
\text{Av} \! \! \sum_{r, s_1,s_2} \! \left| \mathcal{M} \right|^2
&= \frac{512 \, \pi \, g^2 \, e^2 \, \varepsilon^4}{m \, \Gamma(A')}\left[ A - m^2\Dot{k}{k_2}\Dot{k}{k_1}\right] \delta\left( M^2 - 2\Dot{k}{p} - m^2\right) \,,
\label{eq:finalwithspin}
\end{align}
where
\begin{align}
    A & = \frac{1}{4}\left(M^2 - m^2\right)^2 \left(\frac{1}{2}m^2 + m_f^2\right) + \frac{1}{4} m^4 M^2\,.
    \label{eq:Aconstant}
\end{align}
This is our final result for the squared Feynman amplitude {\em with} spin correlations in the narrow-width approximation.

%****************************************
\subsection{Squared Feynman Amplitude without Spin Correlations}
%****************************************

We will now repeat the same calculation, but omitting spin correlations.  We first write the Feynman amplitude \cref{eqn:main-feynman-amplitude} in a way that allows us to interpret it as a superposition of amplitudes for processes involving definite dark photon polarizations. Writing the propagator factor as the sum of dark photon polarization vectors, $- P_{\nu \alpha} = \sum \varepsilon_\nu^* (q) \varepsilon_\alpha (q)$, we can write the amplitude of \cref{eqn:main-feynman-amplitude} as
\begin{align}
    i\mathcal{M} & = \frac{i}{q^2 - m^2 - im \, \Gamma(A')}\sum_l i \varepsilon g M^\nu_r(p, q)\varepsilon_\nu^{*(l)}(q)\cdot \varepsilon^{(l)}_\alpha(q) \left( -i \varepsilon e N^\alpha_{s_1, s_2}(k_1, k_2) \right) \nonumber\\
     & = \frac{i}{q^2 - m^2 - im \, \Gamma(A')}\sum_l\Pi^l_r(p, q)\cdot \Delta^l_{s_1, s_2}(q, k_1, k_2)\,,
\end{align}
where $\Pi^l_r(p, q)$ is the amplitude to produce a photon with momentum $p$ and polarization $r$ and a dark photon with momentum $q$ and polarization $l$ from pion decay, and $\Delta^l_{s_1, s_2}(q, k_1, k_2)$ is the amplitude for a dark photon with momentum $q$ and polarization $l$ to decay to an electron with momentum $k_1$ and polarization $s_1$ and a positron with momentum $k_2$ and polarization $s_2$. 

The squared amplitude is
\begin{align}
\left| \mathcal{M} \right|^2 
&= \frac{1}{\left|q^2 - m^2 - im \, \Gamma(A')\right|^2}\left|\sum_l\Pi^l_r\cdot \Delta^l_{s_1, s_2}\right|^2\,.
\end{align}
To neglect spin correlations, we approximate this expression as
\begin{align}
\left| \mathcal{M} \right|^2 
&\approx \frac{1}{\left|q^2 - m^2 - im \, \Gamma(A')\right|^2}\sum_l\left|\Pi^l_r\right|^2 \cdot \frac{1}{3} \sum_l\left|\Delta^l_{s_1, s_2}\right|^2\,,
\end{align}
where we have averaged over the three intermediate dark photon polarizations. This yields a result similar to \cref{eqn:main-feynman-amplitude-squared-with-spin-correlations}:
\begin{align}\label{eqn:main-feynman-amplitude-squared-without-spin-correlations}
\text{Av} \! \! \sum_{r, s_1,s_2} \! \left| \mathcal{M} \right|^2 
&\approx g^2 \, e^2 \, \varepsilon^4 \, \frac{\frac{1}{3}M^{\nu\nu'}P_{\nu\nu'}P_{\alpha\alpha'}N^{\alpha\alpha'}}{\left|q^2 - m^2 - im \, \Gamma(A')\right|^2} = \frac{128}{3} g^2 \, e^2 \, \varepsilon^4 \frac{\left(q^2 + 2m_f^2\right)\left(M^2 - q^2\right)^2}{\left|q^2 - m^2 - im \, \Gamma(A')\right|^2}\nonumber\\
& = 512 \, g^2 \, e^2 \, \varepsilon^4 \frac{\frac{2}{3}A - \frac{1}{6}M^2q^4}{\left|q^2 - m^2 - im \, \Gamma(A')\right|^2}\,.
\end{align}

Applying the narrow-width approximation, as in \cref{subsec:Feynman amplitude with spin correlation}, we find the averaged squared amplitude
\begin{align}
\text{Av} \! \! \sum_{r, s_1,s_2} \! \left| \mathcal{M} \right|^2
&\approx \frac{512 \, \pi \, g^2 \, e^2 \, \varepsilon^4}{m \, \Gamma(A')}\left[ \frac{2}{3}A - \frac{1}{6}M^2 m^4 \right] \delta\left( M^2 - 2\Dot{k}{p} - m^2\right) \,,
\label{eq:finalwithoutspin}
\end{align}
where $A$ is given in \cref{eq:Aconstant}.  This is our final result for the squared Feynman amplitude {\em without} spin correlations in the narrow-width approximation. 

Comparing \cref{eq:finalwithoutspin} to \cref{eq:finalwithspin}, the final result with spin correlations, the main difference is the lack of the $\Dot{k}{k_1}\Dot{k}{k_2}$ term in \cref{eq:finalwithoutspin}. Without this term, the fermion energy spectrum will be uniform, as will be shown in the next section. A uniform spectrum can be predicted simply by analyzing the $\pi^0\to\gamma A'\to\gamma e^+e^-$ decay as a sequence of classical decays described only by special relativity and 4-momentum conservation, as discussed in \cref{appendix:pdf-calculus-three-body}.

%****************************************
\section{Fermion Energy Distributions}
\label{sec:differential-decay-width-in-the-meson-com}
%****************************************

Given the squared amplitudes of \cref{sec:feynman-amplitudes}, we now calculate the differential decay width, $d\Gamma/dE_1$, where $E_1$ is the energy of the final-state electron in the meson frame.  The energy distributions of the electron and positron are identical.  We will begin with the full calculation with spin correlations in \cref{sec:decay-in-COM-with-spin}.  Because of the similarity of the squared amplitudes with and without spin correlations in \cref{eq:finalwithoutspin,eq:finalwithspin}, respectively, it will then be easy to calculate the result without spin correlations in \cref{sec:decay-in-COM-without-spin}.

%****************************************
\subsection{Fermion Energies with Spin Correlations in the Meson Frame} 
\label{sec:decay-in-COM-with-spin}
%****************************************

The differential decay width with spin correlations in the narrow-width approximation is
\begin{align}
\de^9 \Gamma &= \frac{1}{2M} 
\frac{1}{(2 \pi)^3 2 E} 
\frac{1}{(2 \pi)^3 2 E_1} 
\frac{1}{(2 \pi)^3 2 E_2} 
\text{Av} \! \! \sum_{r, s_1,s_2} \! \left| \mathcal{M} \right|^2 (2\pi)^4 \, \delta^{(4)}(k -k_1 - k_2 - p) \,
\de^3\bm{k}_1\de^3\bm{k}_2\de^3\bm{p} \\
&= \frac{g^2 \, e^2 \, \varepsilon^4}{\pi^4} 
\frac{1}{M m E E_1 E_2 \Gamma(A')}
(A - M^2 m^2 E_1 E_2) \, \delta( M^2 - 2 M E - m^2) \nonumber \\
& \qquad \qquad \times 
\delta(M - E_1 - E_2 - E) \, \delta^{(3)}( \bm{k}_1 + \bm{k}_2 + \bm{p}) \, \de^3\bm{k}_1\de^3\bm{k}_2\de^3\bm{p} \ .
\end{align}
Using the identity $\delta(M^2 - 2M\left|\bm{p}\right| - m^2) = \frac{1}{2M}\delta(E - \left|\bm{p}\right|)$, this can be written as 
\begin{align}
    \de^9\Gamma & = \frac{g^2 \, e^2 \, \varepsilon^4}{2\pi^4} 
\frac{1}{M^2 m E E_1 E_2 \Gamma(A')}
(A - M^2 m^2 E_1 E_2) \, \delta(E - \left|\Vec{p}\right|) \nonumber \\
& \qquad \qquad \times 
\delta(M - E_1 - E_2 - E) \, \delta^{(3)}( \bm{k}_1 + \bm{k}_2 + \bm{p}) \, \de^3\bm{k}_1\de^3\bm{k}_2\de^3\bm{p}\,.
\end{align} 

We now work toward determining the electron spectrum $d\Gamma/d E_1$.  We can easily first integrate with respect to $\bm{p}$, which yields
\begin{align}\label{eq:temp-1}
\de^6\Gamma & = \frac{g^2 \, e^2 \, \varepsilon^4}{2\pi^4} 
\frac{A - M^2 m^2 E_1 E_2}{M^2 m E E_1 E_2 \Gamma(A')}\, \delta(E - \left|\bm{k}_1 + \bm{k}_2\right|)\delta(M - E_1 - E_2 - E) \de^3\bm{k}_1\de^3\bm{k}_2 \, .
\end{align}
We next integrate with respect to $\bm{k}_2$, yielding
\begin{align}
\de^3\Gamma & = \frac{g^2 \, e^2 \, \varepsilon^4}{2\pi^4} 
\frac{A - M^2 m^2 E_1(E' - E_1)}{M^2 m E E_1 (E' - E_1) \Gamma(A')}  \de^3\bm{k}_1 \int_{\bm{k}_2}\delta(E - \left|\bm{k}_1 + \bm{k}_2\right|)\delta(E' - E_1 - E_2) \,\de^3\bm{k}_2 \, .
\label{eq:d3gamma}
\end{align}
The integral in \cref{eq:d3gamma} does not depend on the direction of $\bm{k}_1$, but only on its magnitude, and so 
\begin{align}
& \! \! \! \! \! \! \! \! \int_{\bm{k}_2}\delta(E - \left|\bm{k}_1 + \bm{k}_2\right|)\delta(E' - E_1 - E_2) \, \de^3\bm{k}_2 \\
    & = 2\pi\int_0^\infty k_2^2\de k_2\delta(E' - E_1 - E_2)\int_0^\pi\de\theta\sin\theta\delta\left(E - \sqrt{k_1^2 + 2k_1 k_2 \cos\theta + k_2^2}\right)\\
    & = 2\pi\int_0^\infty k_2^2\de k_2\delta(E' - E_1 - E_2)\int_{-1}^{1}\de u\delta\left(E - \sqrt{k_1^2 + 2k_1 k_2 u + k_2^2}\right) \\
    & = 2\pi\int_0^\infty k_2^2\de k_2\delta(E' - E_1 - E_2)\times\frac{E}{k_1k_2}\Theta\left(k_1 + k_2 - E\right)\Theta\left(E - \left|k_1 - k_2\right|\right)\\
    & = \frac{2\pi E}{k_1}\int_{\left|k_1 - E\right|}^{k_1 + E} k_2\de k_2\delta(E' - E_1 - E_2)\\
    & = \frac{2\pi E\left(E' - E_1\right)}{k_1}\Theta\left(E_1 - E_{\textrm{1min}}\right)\Theta\left(E_{\textrm{1max}} - E_1\right)\,.
\end{align}
Here we have first used spherical symmetry to set $\bm{k}_1 = k_1 \bm{e}_z$ and calculate the integral in spherical coordinates.  Next we used the substitution $u = \cos\theta$. The Heaviside theta functions come from the integral of the delta function $\delta\left(E - \sqrt{k_1^2 + 2k_1 k_2 u + k_2^2}\right) = \frac{E}{k_1 k_2}\delta\left(u - u_0\right)$, where $u_0$ must be confined to $-1 < u_0 < 1$, which is equivalent to $\left|k_1 - k_2\right| < E < k_1 + k_2$. The last step uses the facts that $E_2\de E_2 = k_2\de k_2$ and that the electrons have a minimum and maximum energy, defined by $E' - E_{\textrm{1min/1max}} = E_2 = \sqrt{\left(k_1\pm E\right)^2 + m_f^2}$, that is,
\begin{align}\label{eq:com-spectrum-cutoff}
    E_{\textrm{1max/1min}} = \frac{1}{2}\left(E' \pm E\sqrt{1 - \frac{4m_f^2}{m^2}} \ \right)\,.
\end{align}
These kinematic limits are determined by the fact that, in the pion frame, the electrons will be the most (least) energetic when they are produced parallel (antiparallel) to the direction of the dark photon momentum.  We find, then, that
\begin{align}
\de^3\Gamma & = \frac{g^2 \, e^2 \, \varepsilon^4}{2\pi^4} 
\frac{A - M^2 m^2 E_1(E' - E_1)}{M^2 m E E_1 (E' - E_1) \Gamma(A')} \frac{2\pi E(E' - E_1)}{k_1}\Theta\left(E_1 - E_{\textrm{1min}}\right)\Theta\left(E_{\textrm{1max}} - E_1\right) \de^3\bm{k}_1 \,,
\end{align}
and using $\de^3\bm{k}_1 = 4\pi k_1^2\de k_1 = 4 \pi k_1 E_1 \de{E_1}$, we arrive at our final result
\begin{align}
\frac{\de\Gamma}{\de E_1} & = \frac{4g^2 \, e^2 \, \varepsilon^4}{\pi^2} 
\frac{A - M^2 m^2 E_1(E' - E_1)}{M^2 m\Gamma(A')}\Theta\left(E_1 - E_{\textrm{1min}}\right)\Theta\left(E_{\textrm{1max}} - E_1\right)\,.
\end{align}

The integral of $\frac{\de\Gamma}{\de E_1}$ over the interval between the bounds given in \cref{eq:com-spectrum-cutoff} gives us the total width
\begin{align}\label{eq:total-decay-width}
     \Gamma_{\textrm{tot}}\equiv\Gamma(\pi^0\to\gamma e^-e^+) & = \frac{2g^2\varepsilon^2}{\pi} M^3\left(1 - \frac{m^2}{M^2}\right)^3\times\frac{e^2\varepsilon^2}{12\pi}m\sqrt{1 - \frac{4m_f^2}{m^2}}\left(1 + \frac{2m_f^2}{m^2}\right)\frac{1}{\Gamma\left(A'\right)}\, .
\end{align}
This agrees with the expected result that the total decay width factorizes into production and decay parts given in \cref{eq:mesondecaywidth,eq:darkphotondecaywidth}, that is,
\begin{align}
    \Gamma(\pi^0\to\gamma e^-e^+) = \Gamma(\pi^0 \to \gamma A')\times\frac{\Gamma(A'\to e^+e^-)}{\Gamma\left(A'\right)} \, .
\end{align}
If we normalize the differential decay width by dividing by the total width, we find that the probability distribution for the electron energy $E_1$ is
\begin{align}
    \pdensity_{\textrm{COM}}^{\textrm{with}}(E_1) & \equiv 
    \frac{1}{\Gamma_{\textrm{tot}}} \frac{\de\Gamma}{\de E_1} = \frac{24\left[ A - m^2M^2 E_1\left(E' - E_1\right)\right] }
    {m^2 M^5\sqrt{1 - \frac{4m_f^2}{m^2}}\Bigl[ 1 \! + \! \frac{2m_f^2}{m^2}\Bigr] \! \Bigl[1 \! - \! \frac{m^2}{M^2}\Bigr]^3}
    \Theta \! \left(E_1 \!- \! E_{\textrm{1min}}\right) \Theta \! \left(E_{\textrm{1max}} \!- \! E_1\right) .\label{eq:pdensitywith}
\end{align}

%****************************************
\subsection{Fermion Energies without Spin Correlations in the Meson Frame}
\label{sec:decay-in-COM-without-spin}
%****************************************

Using the techniques discussed above, we can easily find that the differential decay width without spin correlations in the narrow-width approximation is
 \begin{align}
        \frac{\de\Gamma}{\de E_1} & = \frac{4g^2e^2\varepsilon^4}{\pi^2}\frac{1}{mM^2\Gamma(A')}\left(\frac{2}{3}A - \frac{1}{6}M^2m^4\right)\Theta\left(E_1 - E_{\textrm{1min}}\right)\Theta\left(E_{\textrm{1max}} - E_1\right)\nonumber\\
        & = \frac{g^2e^2\varepsilon^4}{3\pi^2}\frac{1}{mM^2\Gamma(A')}\left(m^2 + 2m_f^2\right)\left(M^2 - m^2\right)^2\Theta\left(E_1 - E_{\textrm{1min}}\right)\Theta\left(E_{\textrm{1max}} - E_1\right)\,.\label{eq:differential-decay-width-without-spin-correlation}
    \end{align}
In the pion frame, then, the energy spectrum is flat with sharp cutoffs at $E_{1\textrm{min}}$ and $E_{1\textrm{max}}$, and the electron energy probability distribution is 
\begin{align}
    \pdensity_{\textrm{COM}}^{\textrm{without}}(E_1) & \equiv \frac{1}{\Gamma_{\textrm{tot}}}\frac{\de\Gamma}{\de E_1} = 
        \frac{2M}{M^2 - m^2}\frac{1}{\sqrt{1 - \frac{4m_f^2}{m^2}}}\Theta\left(E_1 - E_{\textrm{1min}}\right)\Theta\left(E_{\textrm{1max}} - E_1\right)\,.\label{eq:pdensitywithout}
\end{align}
Probability calculus is used to show that this result follows from special relativity in \cref{appendix:pdf-calculus-three-body}. 

\begin{figure}[tbp]
\centering
\includegraphics[width=0.49\textwidth]{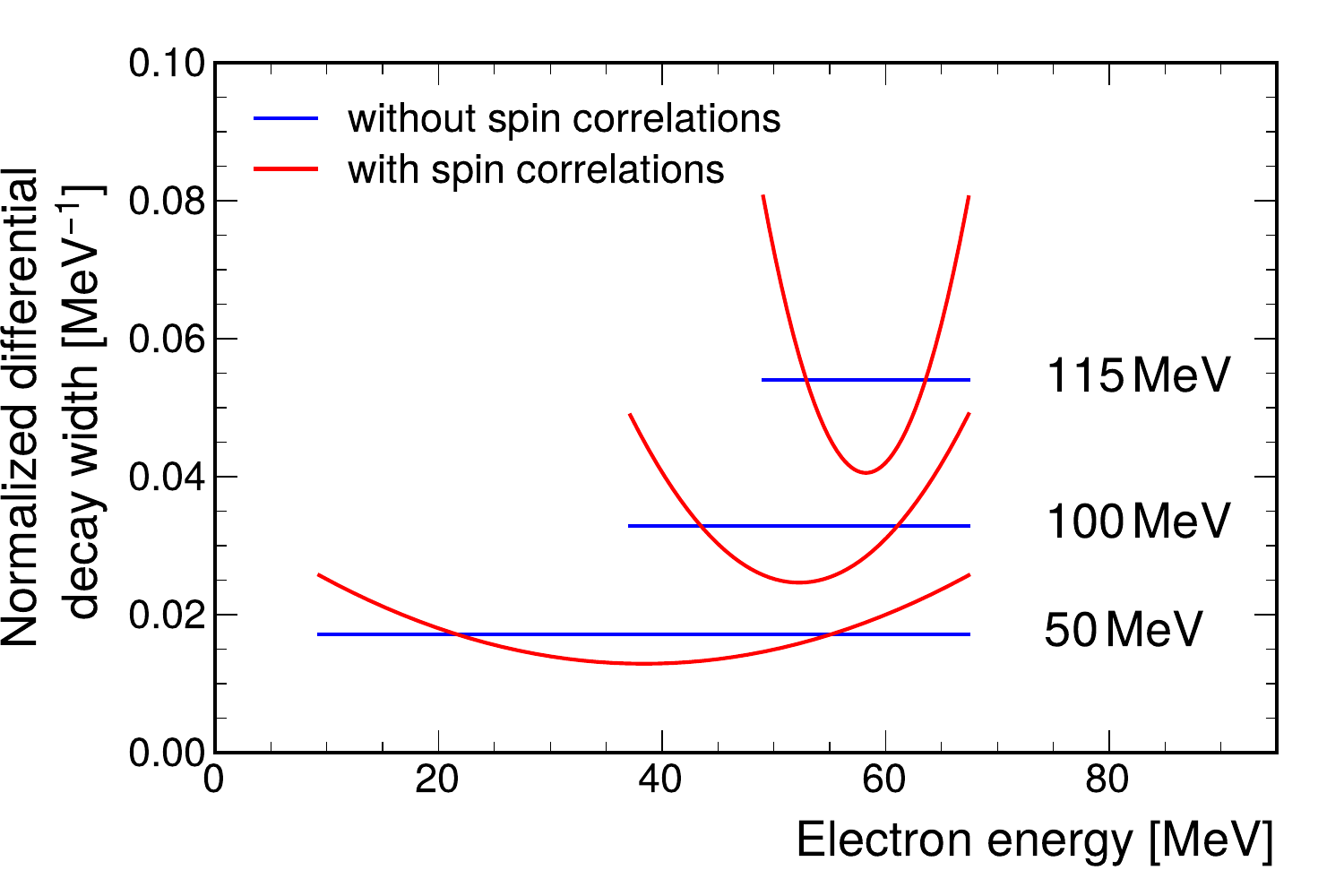}
\hfill
\includegraphics[width=0.49\textwidth]{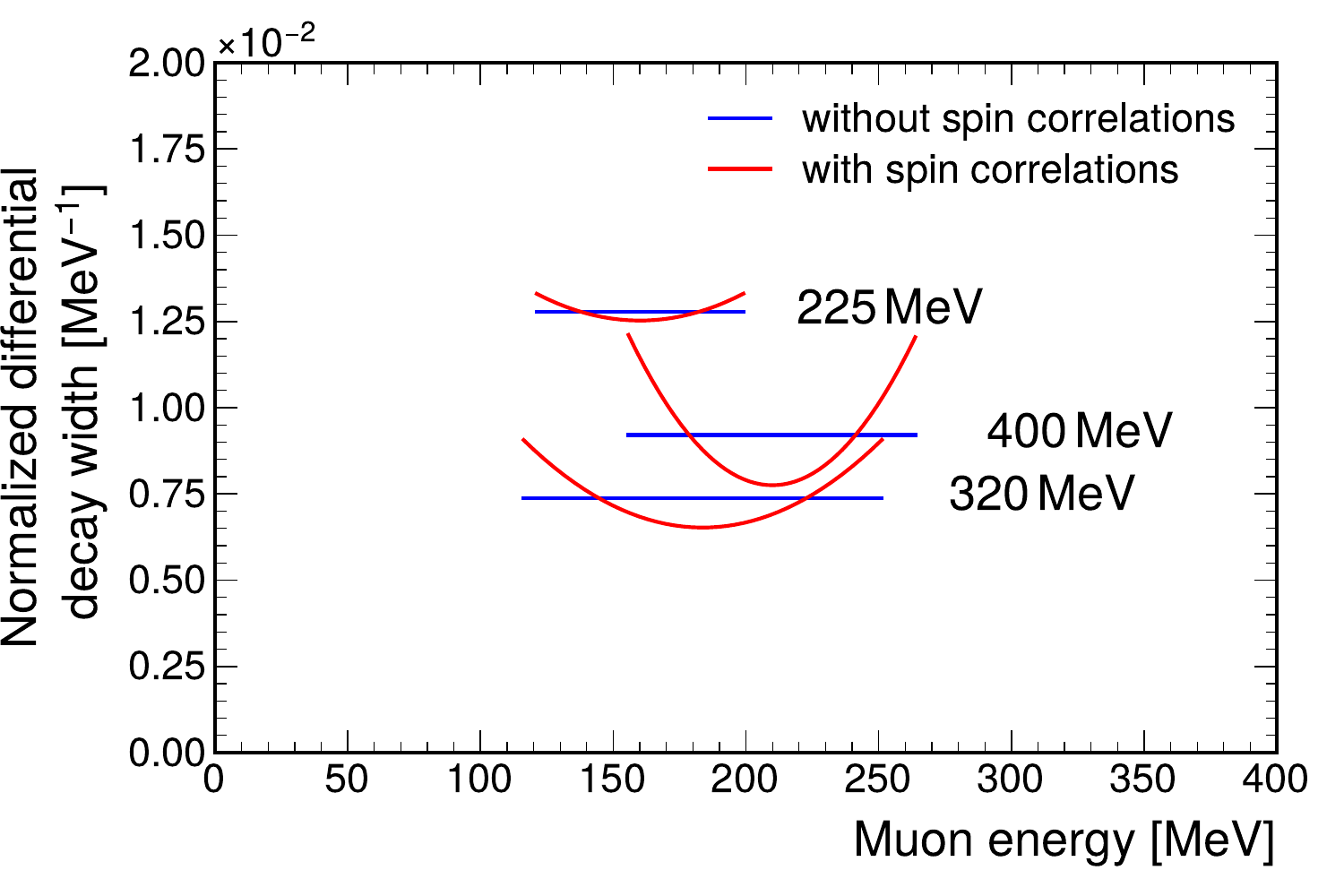}
\caption{The energy probability distributions (normalized differential decay widths) of final-state leptons produced by $\pi^0 \to \gamma A'\to \gamma e^+ e^-$ (left) and $\eta^0 \to \gamma A' \to \gamma \mu^+ \mu^-$ (right) in the meson frame with and without spin correlations for the dark photon masses indicated. }
    \label{fig:COM-pdfs}
\end{figure}

The results of \cref{eq:pdensitywith,eq:pdensitywithout} for the differential decay widths with and without spin correlations, respectively, are shown in \cref{fig:COM-pdfs} for two representative examples.  Without spin correlations, the energy spectrum is flat between the kinematic endpoints. However, with spin correlations, the energy spectrum is enhanced for energies near the endpoints.  The total energy of the $e^+e^-$ pair is, of course, fixed to be the dark photon energy $E' = E_1 + E_2$ of \cref{eq:energies}, but spin correlations enhance the asymmetry between the $e^-$ and $e^+$ energies.  As we see in \cref{fig:COM-pdfs}, for some choices of mass parameters, the resulting enhancement may be significant.  

This result may also be understood in terms of angular correlations.  Without spin correlations, the $e^{\pm}$ are produced isotropically in the dark photon frame.  However, spin correlations enhance the production of $e^{\pm}$ parallel and anti-parallel to the momentum of the dark photon in the pion frame.  When the $e^{\pm}$ are boosted into the pion frame, then, the probabilities that they have energies near their minimal and maximal possible energies are enhanced.  We will return to the impact of spin correlations on angular distributions in more detail in \cref{sec:angulardistributions}.

%****************************************
\subsection{Impact of Spin Correlations on Fermion Energies in the Meson Frame}
%****************************************

To quantify the discrepancy between two energy probability distributions, $\rho_1(E)$ and $\rho_2(E)$, we may define the following discrepancy parameter
\begin{align}
    \Delta & \equiv \frac{1}{2}\int \left|\pdensity_1(E) - \pdensity_2(E) \right| \, \de E\,.
\end{align}
This quantity measures the area between the two probability distributions:~if they are identical, $\Delta = 0$, but if they have no overlap, then $\Delta = 1$. The benefits of using this quantity are that it is easy to calculate, and it is Lorentz invariant, if the integral is taken over the all possible energies. 

For the two meson-frame distributions of \cref{eq:pdensitywith,eq:pdensitywithout}, this discrepancy parameter is 
\begin{align}
    \Delta & = \frac{1}{2}\int \left|\pdensity_{\textrm{COM}}^{\textrm{with}} (E_1) - \pdensity_{\textrm{COM}}^{\textrm{without}}(E_1)\right| \de E_1 \, , 
    \label{eq:DeltaSpecific}
\end{align}
which is identical to the quantity defined in \cref{eq:Delta}.  It has the remarkably simple form
\begin{align}
    \Delta & = \frac{1}{6\sqrt{3}} \ \frac{m^2 - 4m_f^2}{m^2 + 2m_f^2}\leq\frac{1}{6\sqrt{3}}\approx 9.6\%\,.
\label{eq:discrepancy}
\end{align}
We see that, in the limit $2m_f\to m$, the spin correlation effect vanishes; in this limit, the fermions are produced at rest in the dark photon frame, and so their energies are dependent only on the dark photon's energy and not its polarization. In contrast, the effect of spin correlations is maximal for $m_f = 0$, and so for light fermions, like electrons, the effect can be significant. 

In the following section, we will calculate the lab-frame probability distributions with and without spin correlations.  Given that $\Delta$ is Lorentz invariant, it also characterizes the discrepancy of the (much more complicated) lab-frame probability distributions that will be derived below in \cref{eq:probability-density-function-without-spin-correlations-lab-frame,eq:probability-density-function-with-spin-correlations-lab-frame}.

%****************************************
\subsection{Fermion Energies in the Lab Frame}
\label{sec:decay-in-LAB-frame}
%****************************************

In this section, we boost the electron energy spectra from the pion frame, \cref{eq:pdensitywith,eq:pdensitywithout}, into the lab frame. To this end, we can apply the probability density calculus discussed in \cref{appendix:pdf-calculus-1to2-decay}. As in \cref{appendix:pdf-calculus-1to2-decay}, all pion-frame quantities will be labeled with asterisks. Let the pion momentum in the lab frame be $p_\pi$. We want to boost the pion frame into the lab frame by the Lorentz factor $\gamma = \sqrt{1 + \frac{p_\pi^2}{M^2}}$.   Using \cref{appendix:general-boosted-pdf}, we find
\begin{align}
    \pdensity_{\textrm{lab}}(E_1) & 
    %\equiv \frac{1}{\Gamma_{\textrm{tot}}}\frac{\de\Gamma}{\de E_1} 
    = \frac{1}{2\gamma\beta}\int_{E_1^* = \gamma\left(E_1 - \beta\sqrt{E_1^2 - m_f^2}\right)}^{E_1^* = 
 \gamma\left(E_1 + \beta\sqrt{E_1^2 - m_f^2}\right)}\frac{\pdensity_{\textrm{COM}}(E_1^*)\, \de E_1^*}{\sqrt{\left(E_1^*\right)^2 - m_f^2}}
 \label{eq:lab-frame-pdf}\,.
\end{align}
The integration bounds of \cref{eq:lab-frame-pdf} restrict the range of $E_1^*$ to
\begin{align}
    \gamma\left(E_1 - \beta\sqrt{E_1^2 - m_f^2} \ \right) < E_1^* < \gamma\left(E_1 + \beta\sqrt{E_1^2 - m_f^2} \ \right)\,.
\end{align}
At the same time, the function $\pdensity_{\textrm{COM}}(E_1^*)$ is non-vanishing in the region
\begin{align}
    E_{1\rm{min}} < E_1^* < E_{1\rm{max}}\,,
\end{align}
where $E_{1\textrm{min}/1\textrm{max}}$ are given in \cref{eq:com-spectrum-cutoff}. These restrictions compete, and so we may summarize the bounds of integration in \cref{eq:lab-frame-pdf} as
\begin{align}\label{integration-bounds}
    E^*_{1\textrm{max}}(E_1) & = \textrm{Min}\left(E_{1\textrm{max}}, \gamma\left(E_1 + \beta\sqrt{E_1^2 - m_f^2} \ \right)\right)\\
    E^*_{1\textrm{min}}(E_1) & = \textrm{Max}\left(E_{1\textrm{min}}, \gamma\left(E_1 - \beta\sqrt{E_1^2 - m_f^2} \ \right)\right) \ .
\end{align}
The lab frame electron energy spectrum is, therefore,
\begin{align}\label{eq:probability-density-function-with-spin-correlations-lab-frame}
    \pdensity_{\textrm{lab}}^{\textrm{with}}(E_1) & = \left.\frac{12\left[ \! \left( \! A \! + \! \frac{m_f^2 m^2 M^2}{2}\right)\cosh^{-1} \! \left(\frac{x}{m_f}\right) - m_f m^2 M^2\sqrt{\left(\frac{x}{m_f}\right)^2 \! \! - \! 1}\left(E' - \frac{x}{2}\right) \! \right] }{\gamma\beta m^2 M^5\sqrt{1 - \frac{4m_f^2}{m^2}}\left(1 + \frac{2m_f^2}{m^2}\right)\left(1 - \frac{m^2}{M^2}\right)^3}\right|^{x = E^*_{1\textrm{max}}(E_1)}_{x = E^*_{1\textrm{min}}(E_1)}
\end{align}
with spin correlations, and
\begin{align}\label{eq:probability-density-function-without-spin-correlations-lab-frame}
    \pdensity_{\textrm{lab}}^{\textrm{without}}(E_1) & = \left.\frac{1}{2\gamma\beta}\frac{1}{E\sqrt{1 - \frac{4m_f^2}{m^2}}}\cosh^{-1}\left(\frac{x}{m_f}\right)\right|^{x = E^*_{1\textrm{max}}(E_1)}_{x = E^*_{1\textrm{min}}(E_1)}
\end{align}
without spin correlations. 

\begin{figure}[tbp]
\centering
\includegraphics[width=0.49\textwidth]{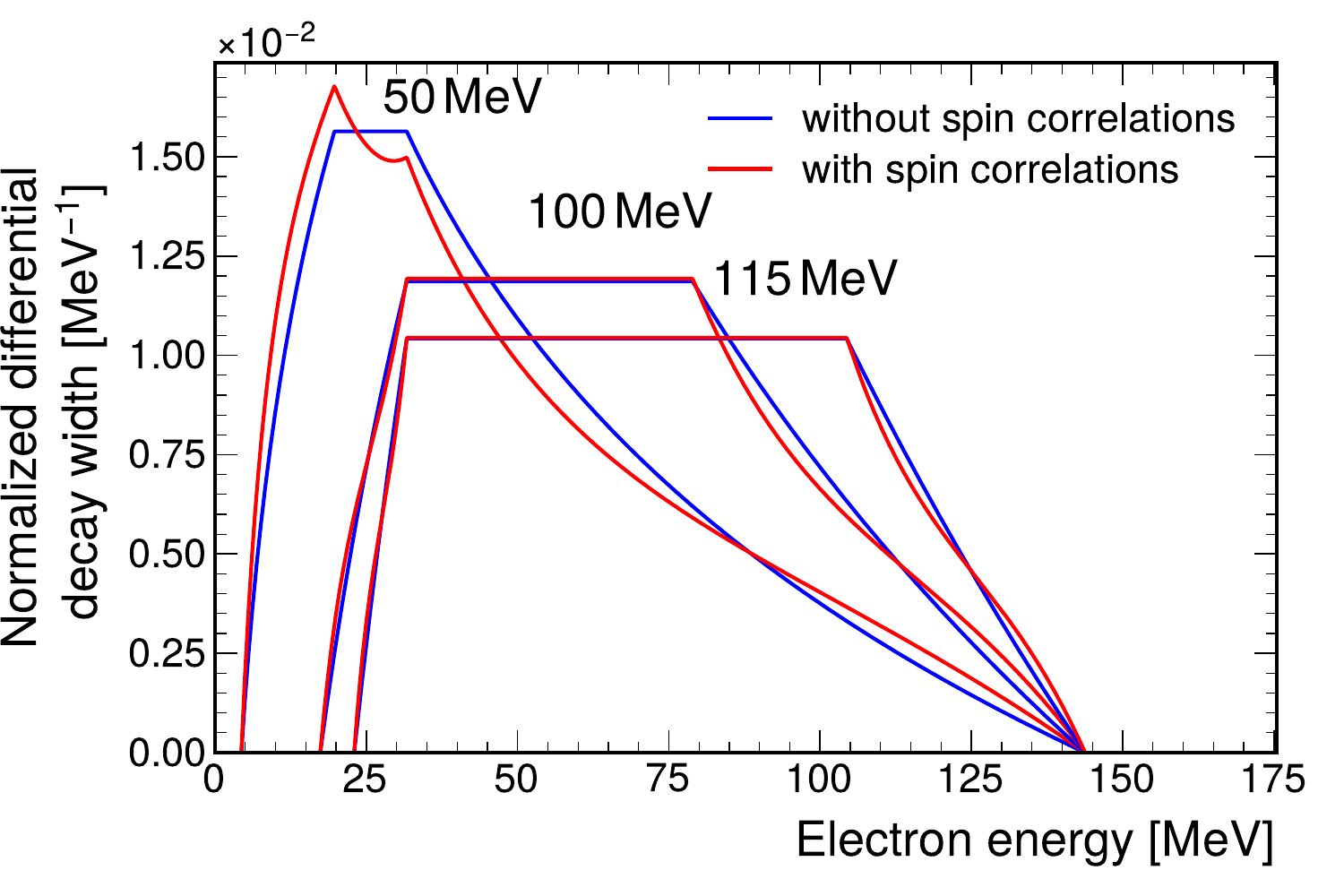}
\hfill
\includegraphics[width=0.49\textwidth]{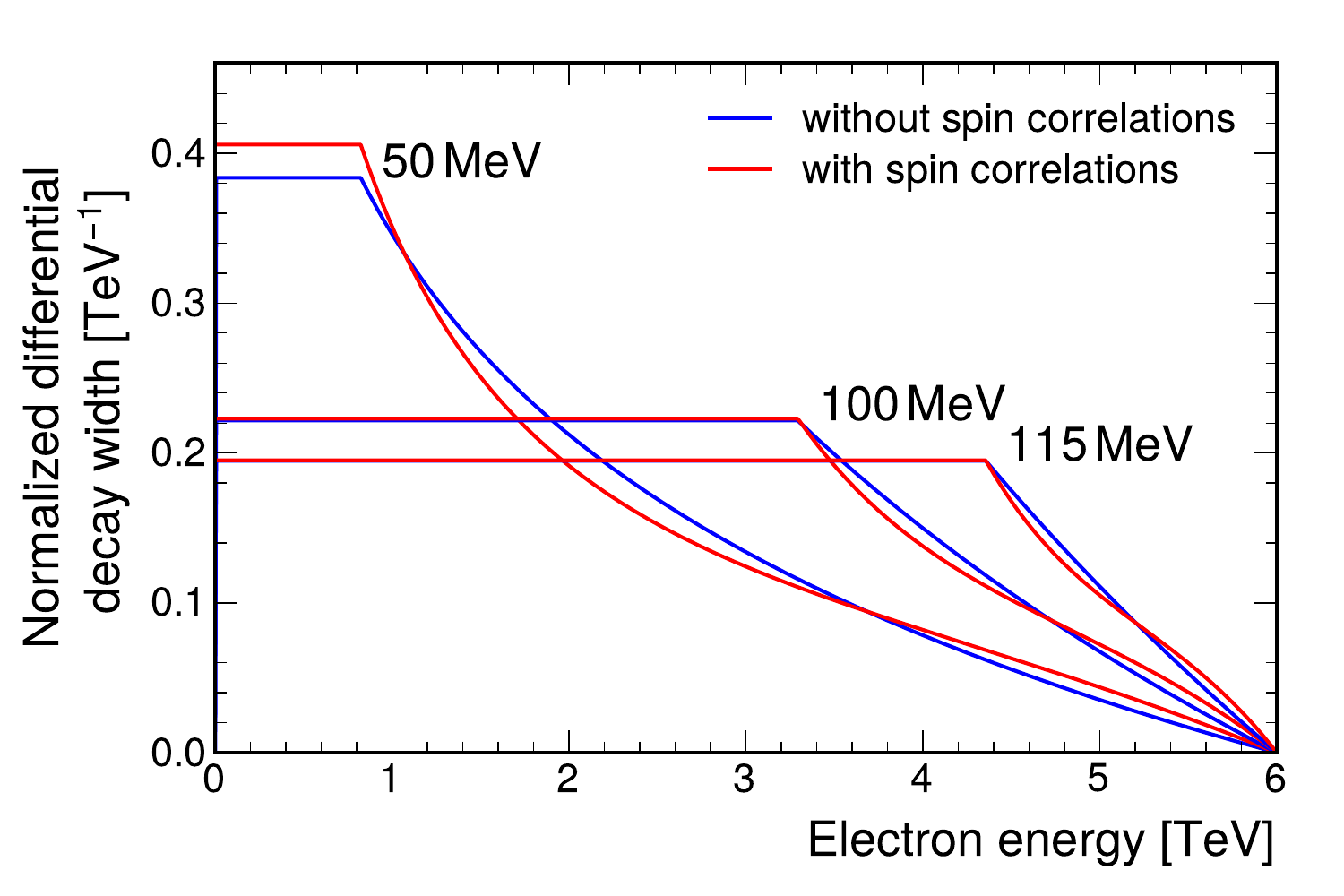}
\caption{The energy probability distributions (normalized differential decay widths) of final-state electrons (or positrons) resulting from $\pi^0 \to \gamma A' \to \gamma e^+ e^-$ in the lab frame for a non-relativistic pion with $p_\pi = 100~\mev$ (left) and an ultra-relativistic pion with $p_\pi =  6~\textrm{TeV}$ (right).  Results are shown with and without spin correlations and for different dark photon masses, as indicated.}
\label{fig:LAB-pdfs-pion}
\end{figure}

\begin{figure}[tbp]
\centering
\includegraphics[width=0.49\textwidth]{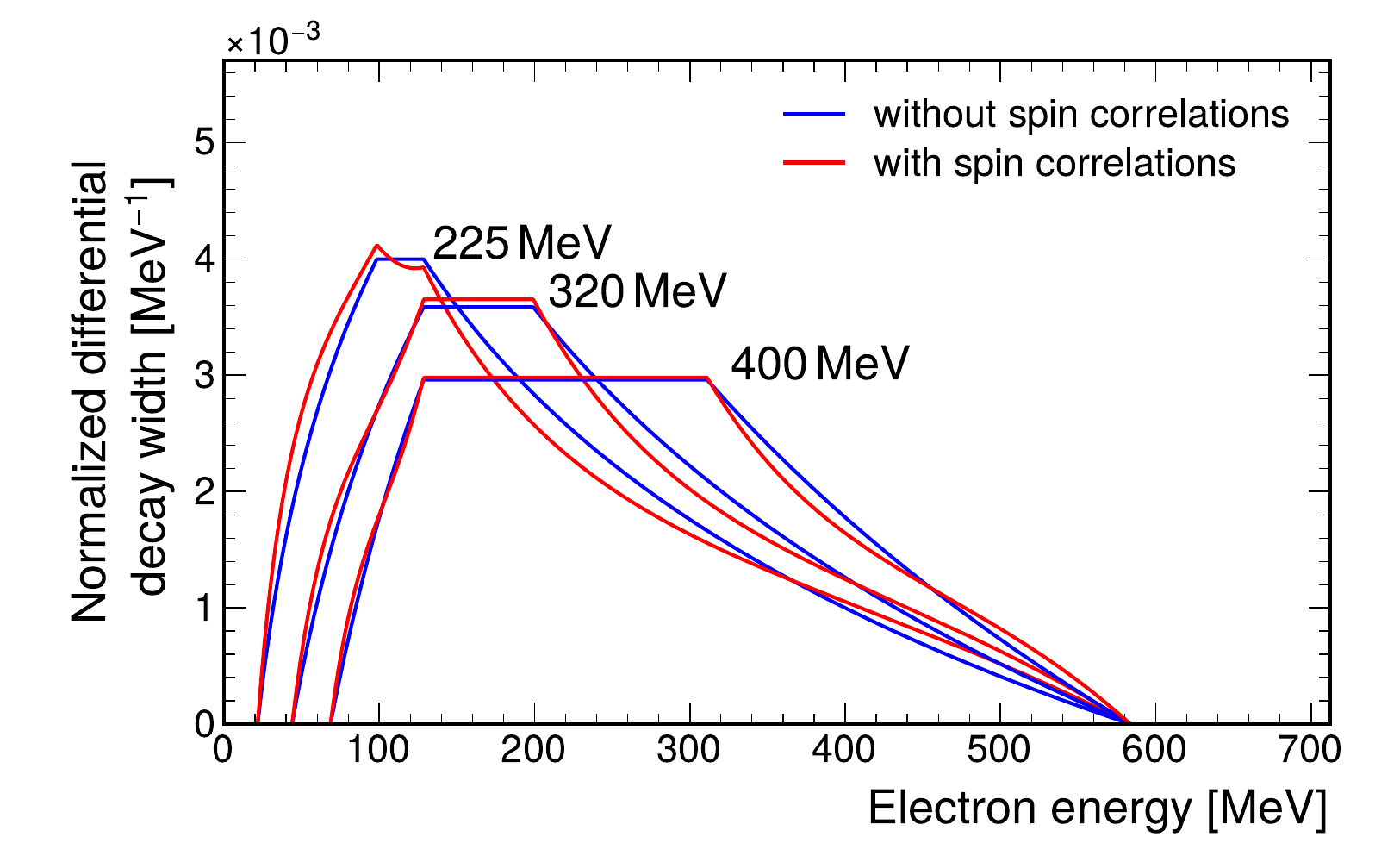}
\hfill
\includegraphics[width=0.49\textwidth]{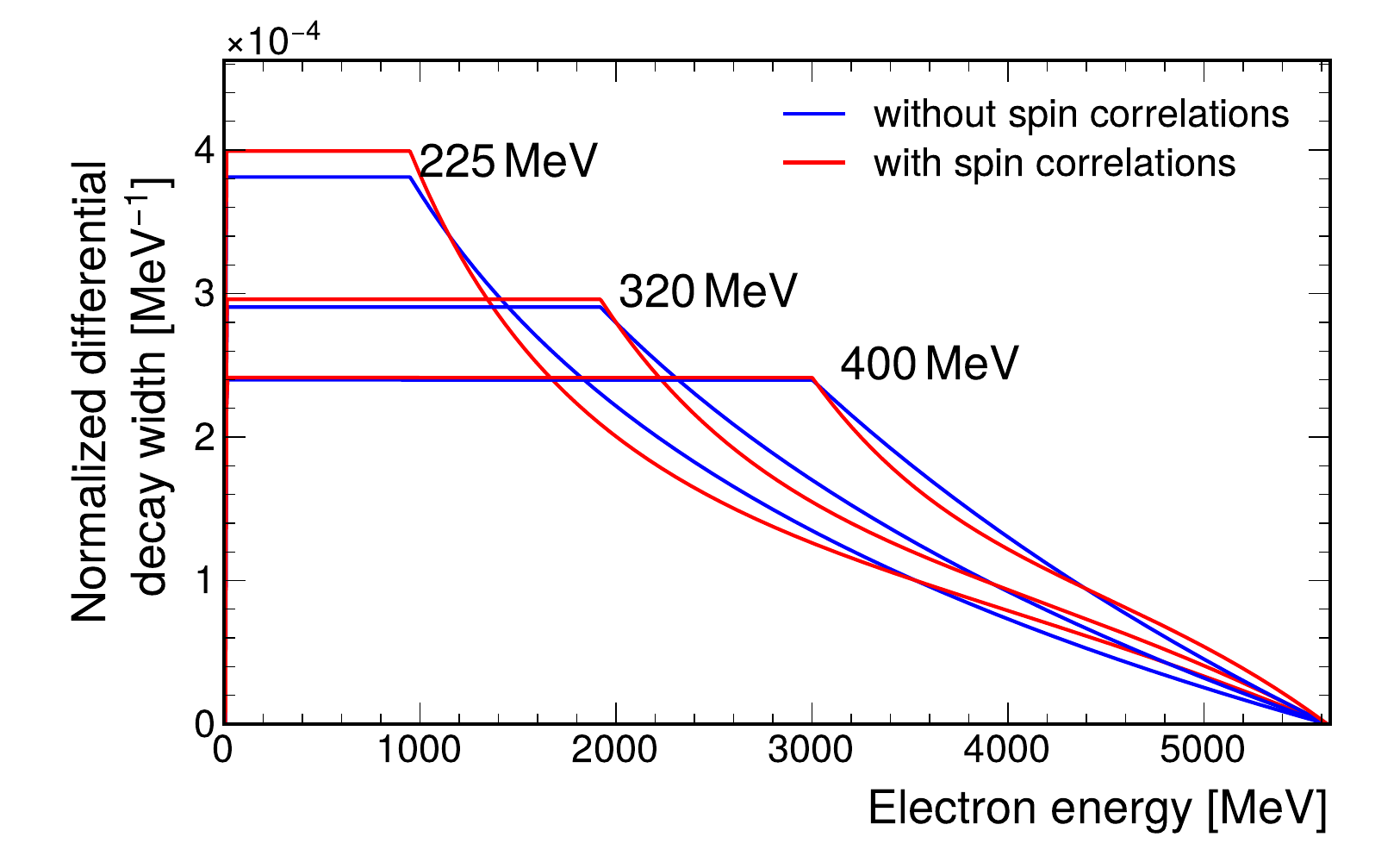}
\vfill
\includegraphics[width=0.49\textwidth]{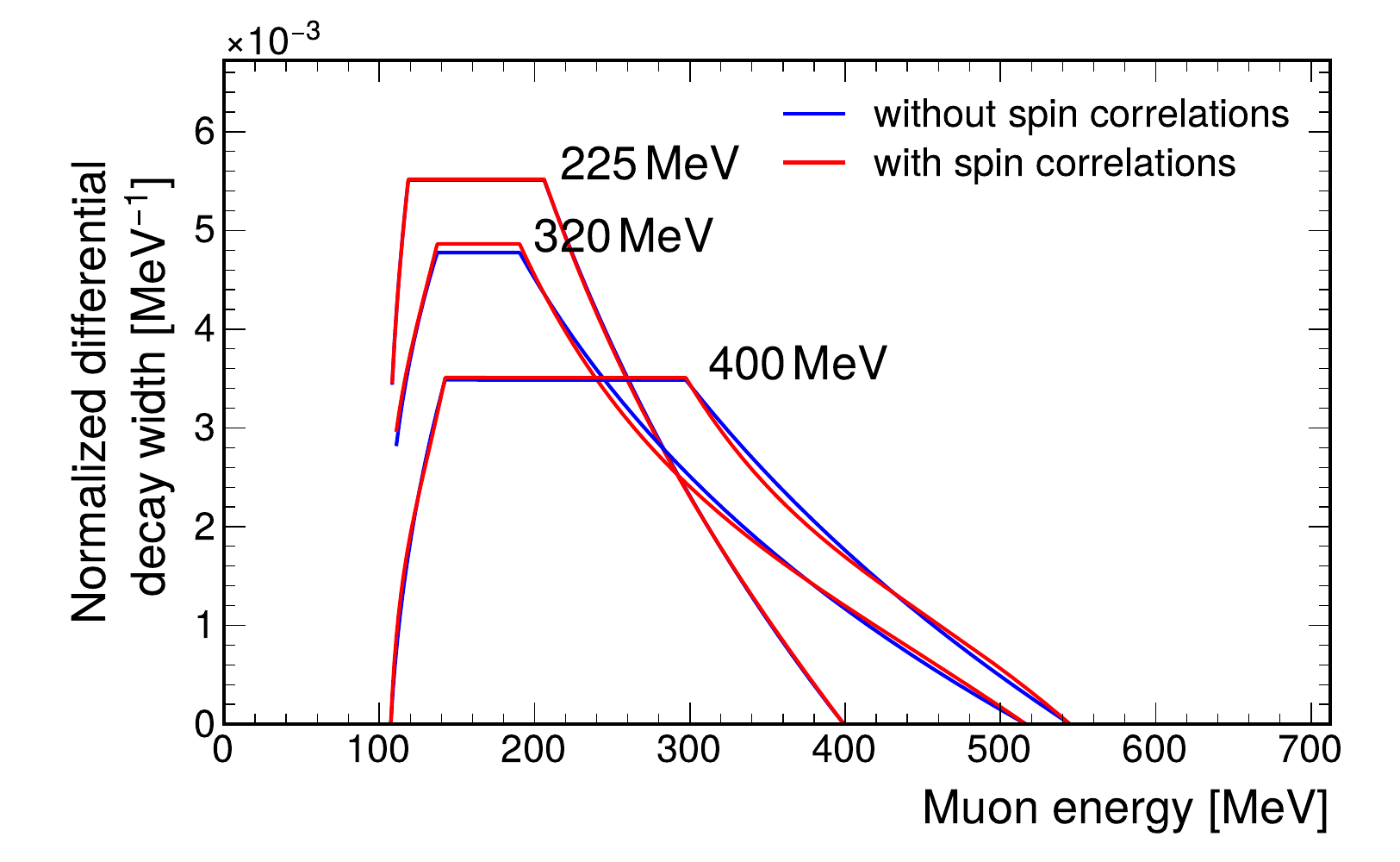}
\hfill
\includegraphics[width=0.49\textwidth]{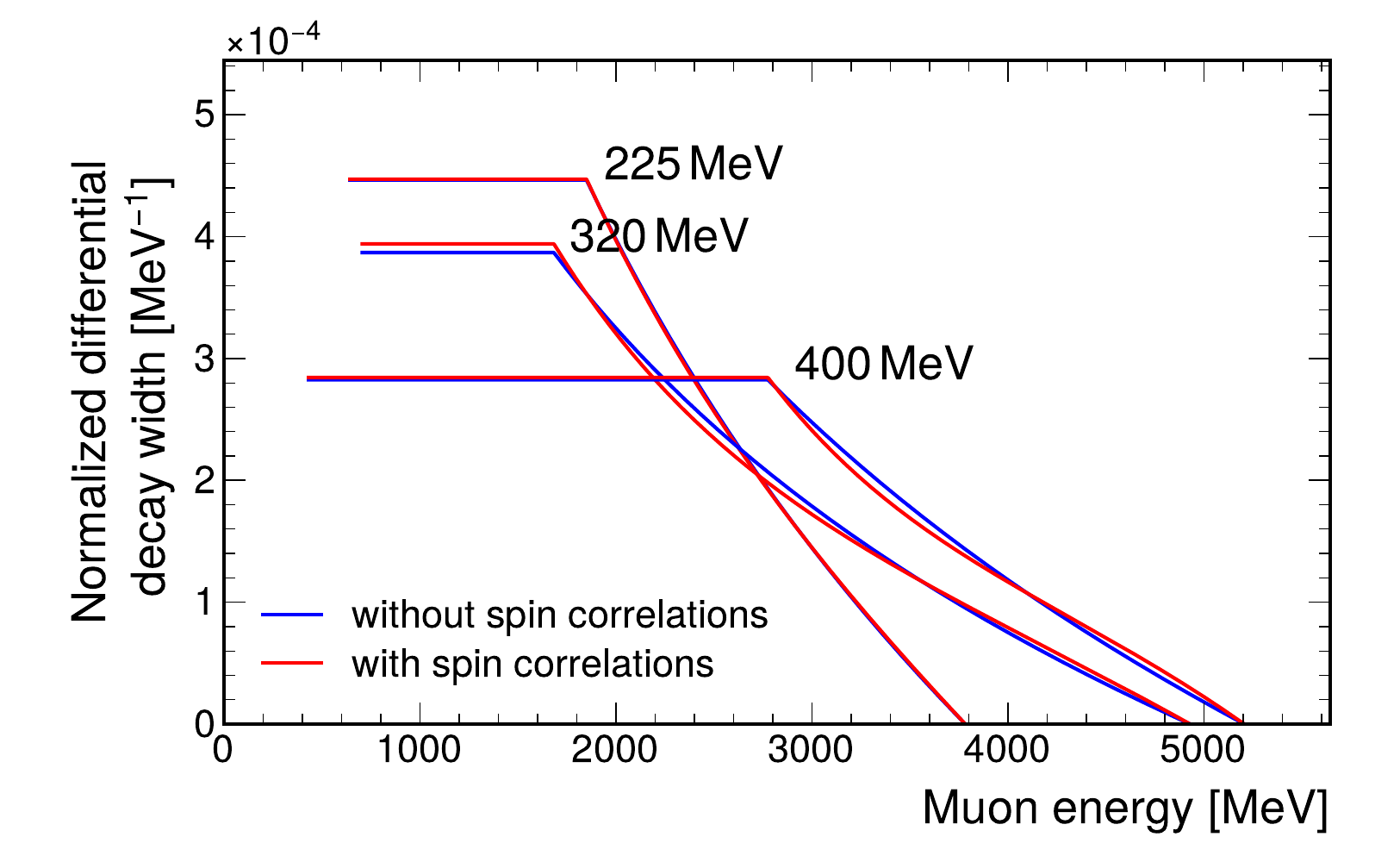}
\caption{The energy probability distributions (normalized differential decay widths) of final-state electrons (top row) or muons (bottom row) resulting from $\eta \to \gamma A' \to \gamma e^+ e^-\,\textrm{or}\,\gamma\mu^+\mu^-$ in the lab frame for a non-relativistic eta with $p_\eta = 500~\mev$ (left column) and a relativistic eta with $p_\eta =  5.6~\textrm{GeV}$ (right column).  Results are shown with and without spin correlations and for different dark photon masses, as indicated. The value $p_\eta =  5.6~\textrm{GeV}$ was chosen to reflect the average $\eta$ momentum in the SHiP experiment.}
\label{fig:LAB-pdfs-eta}
\end{figure}

Examples of these lab-frame energy distributions with and without spin correlations are given in \cref{fig:LAB-pdfs-pion,fig:LAB-pdfs-eta}.  The effect of spin correlations is again evident, although the shapes of the distributions become more intricate.  In particular, the boosted distributions typically have flat plateaus over certain energy ranges, as explained in \cref{appendix:why-is-there-a-plateau?}.  

%****************************************
\section{Fermion Angular Distributions}
\label{sec:angulardistributions}
%****************************************

In addition to modifying the energy distributions of fermions produced in dark photon decays, spin correlations change the angular distributions of these fermions.  This can modify the number of events that pass cuts by changing the number of fermions that travel through a detector, track separation efficiencies, and other experimental observables.  Here we analyze the impact of spin correlations on angular distributions in the dark photon, meson, and lab frames in turn.

%****************************************
\subsection{Fermion Angles in the Dark Photon Frame}
\label{sec:angulardistributionsDarkPhoton}
%****************************************

If spin correlations are neglected, dark photons are considered to be produced unpolarized, and so they decay to fermions isotropically in the dark photon frame.  In this subsection, we show that spin correlations can be included in a semi-classical approach, in which the fermions are produced with a specific, anisotropic distribution in the dark photon frame, which, when boosted, gives the fermions the correct energy distributions derived above in \cref{sec:decay-in-COM-with-spin,sec:decay-in-LAB-frame} when spin correlations are included.  

The energy distribution of fermions in the meson frame with spin correlations is given in \cref{eq:pdensitywith}.  This can be reproduced by decaying the dark photon into fermions in a specific anisotropic distribution.  We can determine this distribution as follows.  Following the notation of \cref{appendix:pdf-calculus-three-body}, we assume the electron's 4-momentum in the dark photon frame is $(E_1^*, k_1^* \bm{n}^*(\theta^*, \phi^*))$, where the asterisk superscripts denote quantities in the dark photon frame, and $\theta^*$ and $\phi^*$ parametrize the $\bm{n}^*$ unit vector's direction in spherical coordinates.  Assume further that the dark photon's 4-momentum in the meson frame is $(E', q \bm{n}(\theta, \phi))$.  The electron's energy in the meson frame is, then, 
\begin{align}
    E_1 & = \gamma\left(E_1^* - \frac{qk_1^*}{\gamma m} \ \bm{n}\left(\theta, \phi\right)\cdot\bm{n}^*\left(\theta^*, \phi^*\right)\right) = \frac{1}{2}\left(E' - E\sqrt{1 - \frac{4m_f^2}{m^2}} \ \cos\theta^*\right)\,,
\end{align}
where $\gamma = E'/m$, and we have assumed the dark photon's 3-momentum to be in the $z$-direction.

\begin{figure}[tbp]
\centering
\includegraphics[width=0.75\textwidth]{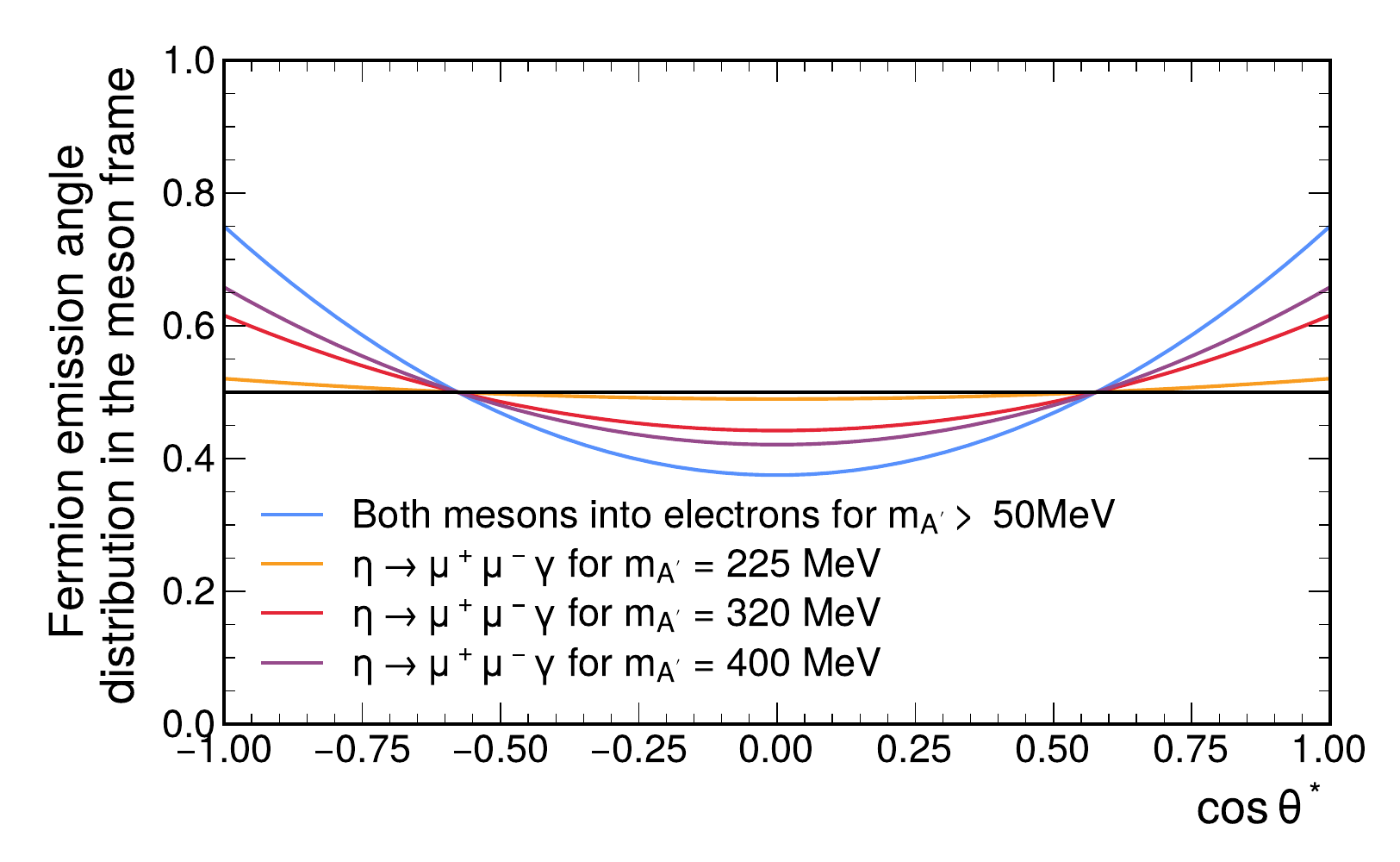}
\caption{Fermion emission angle distributions \cref{eq:angle-distribution-in-dark-photon-frame} without spin correlations (black line) and with spin correlations (colored curves). }
\label{fig:COM-opening-angle}
\end{figure}

Let the probability distribution for the electron emission angle in the meson frame, $\theta^*$, be $\pdensity_{\cos\Theta^*}\left(\cos\theta^*\right)$. To find the angular distribution that leads to the energy distribution of \cref{eq:pdensitywith}, we use \cref{appendix:dependent-pdf-formula-little} in the form
\begin{align}
\rho^{\textrm{with}}_\textrm{COM}\left(E_1\right)\de E_1 & = \pdensity_{\cos\Theta^*}\left(\cos\theta^*\right) \ \de \cos\theta^*\,.
\end{align}
Using the fact that $\de E_1 = \frac{E}{2}\sqrt{1 - \frac{4m_f^2}{m^2}} \ \de \cos\theta^*$, we find
\begin{align}\label{eq:angle-distribution-in-dark-photon-frame}
\pdensity_{\cos\Theta^*}\left(\cos\theta^*\right) = \frac{3}{8}\frac{m^2 + 4m_f^2 + \left(m^2 - 4m_f^2\right)\cos^2\theta^*}{m^2 + 2m_f^2}\,.
\end{align}
In the $m_f \to m/2$ limit, $\pdensity_{\cos\Theta^*}\left(\cos\theta^*\right) \to \frac{1}{2}$; that is, the distribution becomes isotropic, and spin correlations disappear. In the opposite limit $m_f \to 0$, the spin correlations become maximal.  Note also that the angular distribution of \cref{eq:angle-distribution-in-dark-photon-frame} is convex in $\cos\theta^*$, meaning that the fermion is more likely to be emitted parallel ($\cos\theta^* = 1$) or anti-parallel ($\cos\theta^* = -1$) to the dark photon direction in the meson frame than it is to be emitted in a perpendicular direction ($\cos\theta^* = 0$).  This is consistent with the fact that spin correlations increase the asymmetry of the fermion--anti-fermion energy distributions.  Examples of the angular distributions $\pdensity_{\cos\Theta^*}\left(\cos\theta^*\right)$ with and without spin correlations are shown in \cref{fig:COM-opening-angle}.

\cref{eq:angle-distribution-in-dark-photon-frame} is an extremely useful result, because it allows spin correlations to be included in dark photon signal simulators in a simple way.  In such simulators, the production and decay are typically treated separately, as is justified if spin correlations are neglected.  The results derived here show how spin correlations, despite the fact that they result from the interplay between production and decay, can be incorporated solely by modifying the way in which dark photons are decayed, without modifying how they are produced.  We will demonstrate this below in \cref{sec:the-impact-of-spin-correlations}, when we include spin correlations in some common signal simulation frameworks.

%**********************************************
\subsection{Fermion Opening Angles in the Meson Frame}
\label{sec:angles-in-meson-frame}
%**********************************************

In this section, we compute the probability distribution of fermion opening angles $\openingAngle^*$, that is, the angle between the fermion and anti-fermion directions in the meson frame.  (We will use $\openingAngle^*$ to denote opening angles in the meson frame, and in the next subsection, $\openingAngle$ to denote opening angles in the lab frame.)  It is easy to relate $\openingAngle^*$ to the fermion energy in the meson frame.  Using the notation of \cref{fig:pion-decay}, 4-momentum conservation in the form $q^2 = (k_1 + k_2)^2$ implies
\begin{align}
\label{eq:temp-5}
\zee^*\equiv\cos\openingAngle^* = \frac{E_1 E_2 + m_f^2 - \frac{1}{2} m^2}{k_1 k_2} = \frac{E_1\left(E' - E_1\right) + m_f^2 - \frac{1}{2} m^2}{\sqrt{E_1^2 - m_f^2}\sqrt{\left(E' - E_1\right)^2 - m_f^2}}\,.
\end{align}
In the third equality, we used the fact that fermion energies sum up to $E_1 + E_2 = E'$. This also means that $\cos\openingAngle^*$ is symmetric around the value $E_1 = \frac{1}{2}E'$. By explicit computation, we find that the value of $\cos\openingAngle^*$ at the edge of the fermion energy spectrum is
\begin{align}
\left.\cos\openingAngle^*\right|_{E_1 = E_{\textrm{1min}}} & = \textrm{sign}\left(m_f^2\left(M^2 + m^2\right)^2 - M^2 m^4\right) = \textrm{sign}\left(E' - \frac{m^2}{2m_f}\right)\,.    
\end{align}
For fermion masses smaller (larger) than $\frac{m^2}{2E'}$, then, $\cos\openingAngle^*$ reaches a minimum (maximum) at the edge of the spectrum, and at the midpoint, $E_1 = \frac{1}{2}E'$, $\cos\openingAngle^*$ reaches its maximum (minimum). 

The opening angle distribution $\pdensity_{\cos\openingAngleCapital^*}$ can be found with the use of \cref{appendix:dependent-pdf-formula-big}. To proceed, we work in the massless fermion limit, $m_f = 0$. In this limit, \cref{appendix:dependent-pdf-formula-big} takes a simplified form, since $\cos\openingAngle^*$ is a concave function of $E_1$, and so 
\begin{align}
    & \rho_{\cos\openingAngleCapital^*}\left(\zee^*\right) = \int\rho\left(E_1\right)\delta\left(\zee^* - \left(1 - \frac{m^2}{2E_1\left(E' - E_1\right)}\right)\right)\de E_1\nonumber\\
    & = \frac{m^2}{\left(1 - \zee^* \right)^2\sqrt{E'^2 - \frac{2m^2}{1 - \zee^*}}}\left[\rho\left(\frac{1}{2}E' + \frac{1}{2}\sqrt{E'^2 - \frac{2m^2}{1 - \zee^*}} \, \right) + \rho\left(\frac{1}{2}E' - \frac{1}{2}\sqrt{E'^2 - \frac{2m^2}{1 - \zee^*}} \ \right)\right]
    \label{appendix:general-angular-pdf}\\
    & = \frac{2m^2}{\left(1 - \zee^*\right)^2\sqrt{E'^2 - \frac{2m^2}{1 - \zee^*}}}\rho\left(\frac{1}{2}E' + \frac{1}{2}\sqrt{E'^2 - \frac{2m^2}{1-\zee^*}} \ \right)
    \label{appendix:specific-angular-pdf}\, ,
\end{align}
where, to derive \cref{appendix:specific-angular-pdf} from \cref{appendix:general-angular-pdf}, we use the fact that the meson-frame electron energy spectrum is symmetric about $E_1 = \frac{1}{2}E'$, with or without spin correlations.

%****************************************
\subsection{Fermion Opening Angles in the Lab Frame}
\label{sec:angles-in-LAB-frame}
%****************************************

In this section, we aim to gain a general understanding of the impact of spin correlations on opening angles in the lab frame. To do this, we will compute the opening angle in the lab frame, $\psi$, as a function of the electron emission angle with respect to the $z$-axis, $\theta^*$, in the ultra-relativistic limit, where $\gamma = \sqrt{\frac{p_\pi^2}{M^2} + 1}\to\infty$. The $z$-axis is chosen to align with the dark photon momentum in the pion frame, $\bm{q}$, in the same convention used in \cref{appendix:pdf-calculus-three-body}. In the ultra-relativistic limit, we can treat the dark photon momentum in the pion frame as negligible compared to $p_\pi$. In turn, the fermion momentum in the pion frame is approximately the same as the fermion momentum in the dark photon frame, $k_1^*$, from \cref{eq:temp-6}.  The vector $\bm{q}$ divides the opening angle into angles $\psi_-$ and $\psi_+$, and so we may calculate the opening angle as
\begin{align}
    \openingAngle = \psi_+ + \psi_- \ ,
\end{align}
where
\begin{align}
\tan\psi_\pm = \frac{k_1^*\sin\theta^*}{\gamma\left(E_1^* \pm \beta k_1^*\cos\theta^*\right)}
\end{align}
is the ratio of the transverse to longitudinal momenta for the fermion and anti-fermion. The minus sign follows from the fact that the fermion and anti-fermion have opposite momenta in the meson frame. In the ultra-relativistic limit, $\gamma\to\infty$, the opening angle becomes
\begin{align}
\openingAngle\left(\theta^*\right) & = \frac{\sqrt{1 - z^2}}{\gamma}\frac{m\sqrt{\frac{m^2}{4} - m_f^2}}{m_f^2 + \left(\frac{m^2}{4} - m_f^2\right)\left(1 - z^2\right)} \ ,
\label{appendix:opening-angle-ultrarelativistic-limit}
\end{align}
where $z = \cos \theta^*$.  To compute the opening angle probability distribution, $\pdensity_\openingAngleCapital$, we apply \cref{appendix:dependent-pdf-formula} to find
\begin{align}
\rho_{\openingAngleCapital}\left(\openingAngle\right) & = \int_0^1\frac{\de z}{\sqrt{1 - z^2}}\rho_{\cos\Theta^*}\left(z\right)\delta\left(\openingAngle - \frac{\sqrt{1 - z^2}}{\gamma}\frac{m\sqrt{\frac{m^2}{4} - m_f^2}}{m_f^2 + \left(\frac{m^2}{4} - m_f^2\right)\left(1 - z^2\right)}{}\right)\\
    & = \frac{m \sqrt{\frac{m^2}{4} - m_f^2}x_0^2\rho_{\cos\Theta^*}(u_0)}{\gamma\openingAngle^2 u_0\left(m_f^2 - \left(\frac{m^2}{4} - m_f^2\right) x_0^2\right)}\Theta\left(1 - u_0\right)\,,\label{eq:LAB-angular-pdf}
\end{align}
where 
\begin{align}
u_0 & = \sqrt{1 - x_0^2} \\
x_0 & = \frac{m}{2\openingAngle\gamma\sqrt{\frac{m^2}{4} - m_f^2}}\sqrt{1 - \left(\frac{2m_f\openingAngle\gamma}{m}\right)^2} \ .
\end{align}

As a final result, we can calculate the average value of the opening angle with spin correlations, $\braket{\openingAngleCapital}_{\textrm{with}}$, and without spin correlations, $\braket{\openingAngleCapital}_{\textrm{without}}$. Using the opening angle formula \cref{appendix:opening-angle-ultrarelativistic-limit} and emission angle probability density \cref{eq:angle-distribution-in-dark-photon-frame}, we get
\begin{align}
    \braket{\openingAngleCapital}_{\textrm{with}} & = \int_{\theta^*\,=\,0}^{\theta^*\,=\,\pi} \pdensity_{\Theta^*}(\theta^*)\de \theta^*\openingAngle(\theta^*) = \int_{z\,=\,-1}^{z\,=\,1}\frac{ \pdensity_{\cos\Theta^*}(z)\de z}{\sqrt{1 - z^2}}\openingAngle(z)\\
    & = \frac{3}{\gamma}\coth^{-1}\left(\frac{m}{\sqrt{m^2 - 4m_f^2}}\right) - \frac{3}{2\gamma}\frac{m\sqrt{m^2 - 4m_f^2}}{m^2 + 2m_f^2}\,.
\end{align}
The average opening angle without spin correlations is determined from the uniform distribution for the $\cos\Theta^*$ which is $\pdensity_{\Theta^*}\left(\theta^*\right) = \frac{1}{2}$. Explicit calculation gives
\begin{align}
    \braket{\openingAngleCapital}_{\textrm{without}} & = \int_{z\,=\,-1}^{z\,=\,1}\frac{1}{2}\frac{\de z}{\sqrt{1 - z^2}}\openingAngle(z) = \frac{2}{\gamma}\coth^{-1}\left(\frac{m}{\sqrt{m^2 - 4m_f^2}}\right)\,.
\end{align}
Although these results might appear different in the no spin correlation limit, $m\to 2m_f$, they are actually the same. Substituting $m = 2m_f\sqrt{1 + a^2}$, where $a > 0$ and Taylor expanding for small values of $a$, we see that
\begin{align}\label{eq:opening-angle-enhancement}
\braket{\openingAngleCapital}_{\textrm{with}} - \braket{\openingAngleCapital}_{\textrm{without}} & = \frac{4a^5}{45\gamma} + {\cal O}(a^7) \ .
\end{align}
This shows that the no spin correlation limit is correctly recovered as $a\to 0$. By inspection we see that $\braket{\openingAngleCapital}_{\textrm{with}} > \braket{\openingAngleCapital}_{\textrm{without}}$, that is, the average angle between the fermions is always larger if we include spin correlations, which is consistent with the results of simulations that we will show below in \cref{fig:faser_open_angle,fig:ship_open_angle}.

%****************************************
\section{The Impact of Spin Correlations in Representative Experimental Analyses}
\label{sec:the-impact-of-spin-correlations}
%****************************************

In evaluating the effect of spin correlations analytically, we found that the discrepancy parameter $\Delta$ of \cref{eq:discrepancy} is never very big.  Even for the case of final-state electrons, where $m_f \ll m$, which maximizes the spin correlation effect, $\Delta$ is still less than $9.6\%$. However, in a realistic experimental analysis, there will be cuts on the energies of the final state particles, requirements that the particles pass through the decay volume of the detector, and detector effects, all of which can either enhance or suppress the  effect of spin correlations.  These are not accounted for in the purely analytical results we have derived above. We note that there are other theoretical uncertainties in predicting dark photon signal rates, notably, the uncertainty associated with forward hadron production.  To predict the meson flux, different event generators use different hadronization models, leading to a spread in their predictions of the order of $\sim 10\%$, as we discuss below.  As more data are collected in the forward region, this uncertainty may be reduced over time. Of course, the spin correlation effect will be present regardless of which event generator is used.  

In this section, we will explore the effect of experimental cuts and acceptances on the size of spin correlation effects.  It is beyond the scope of this study to carry out detailed detector simulations or to consider all of the many different experimental settings in which dark photon searches have been carried out. Instead, we will explore the impact of spin correlations with truth-level simulations in two representative experimental settings:~a completed search with FASER at the LHC and a proposed search at SHiP, a fixed target experiment.

%****************************************
\subsection{Dark Photons at FASER}
\label{subsec:dark-photons-FASER}
%****************************************

FASER~\cite{Feng:2017uoz,FASER:2018bac,FASER:2022hcn} has searched for dark photons at the LHC.  Proton-proton collisions at the center-of-mass energy of 13.6 TeV at the ATLAS interaction point can produce dark photons through meson decay or dark bremsstrahlung.  These dark photons may then travel approximately 480 m and then decay in the FASER detector.  FASER's current bound~\cite{FASER:2023tle}, based on an integrated luminosity of $27$~fb$^{-1}$ collected in 2022, is the most stringent constraint on dark photons with masses $\sim 17~\text{MeV}-70~\text{MeV}$ and couplings $\varepsilon \sim  2\times 10^{-5} - 1\times 10^{-4}$.  FASER will continue collecting data through the end of LHC Run 3 in 2026, with an expected total integrated luminosity of $300~\text{fb}^{-1}$, and will then continue running in LHC Run 4 from 2030-33~\cite{Boyd:2882503} into the High-Luminosity LHC era, which has an expected total integrated luminosity of $3~\text{ab}^{-1}$.  It is therefore of interest to include experimental cuts and determine whether spin correlations affect the published FASER results~\cite{FASER:2023tle} or might need to be included in future searches based on much larger luminosities.  

To investigate the effect of spin correlations after experimental cuts are imposed, we simulate dark photon decays in FASER with and without spin correlations using Independent Signal Generator (ISG)~\cite{Kling:2845232}, a program that simulates dark photon signals at FASER, which was written to validate FORESEE~\cite{Kling:2021fwx}. ISG is described in detail in Ref.~\cite{Kling:2845232}.  We use ISG with the EPOS-LHC~\cite{Pierog:2013ria} $\pi^0$ spectrum implemented in FORESEE, and the detector is centered 6.5 cm above the beam collision axis or line-of-sight (LOS) to account for the beam half-crossing angle, as described in Ref.~\cite{Kling:2845232}.

Following Ref.~\cite{Kling:2845232}, EPOS-LHC was chosen to generate the $\pi^0$ spectrum, as it gives the middle prediction of the dark photon production rate among the following event generators: EPOS-LHC, SYBILL, and QGSJET. According to Fig.~12 of Ref.~\cite{Kling:2845232}, the dark photon production rate spread between EPOS-LHC and the other two event generators is of the order of 20\%. According to Ref.~\cite{FASER:2024ykc} (Fig.~2, top right corner), EPOS-LHC, SIBYLL, QGSJET and PYTHIAforward have a spread of the order of $10-20\%$ for $\eta$ mesons reconstructed via $\eta\to\gamma\gamma$ decay. The same reference shows that the uncertainty in experimental measurements of the $\eta$ flux is of the order of 10\%, which can be expected to be reduced as more data in the far-forward region becomes available. Although this underlying uncertainty affects the absolute meson yield, the subsequent spin correlation effects, which can reach 10\%, are a consistent feature of the $A'$ production and decay chain, independent of the specific generator used.
The dark photon events are simulated as follows:
\begin{itemize}
\setlength\itemsep{-0.05in}
\item \textit{Without} spin correlations: The dark photons that decay in the decay volume are decayed to $e^+ e^-$ isotropically in the $A'$ rest frame, and the $e^-$ and $e^+$ are then boosted to the lab frame. This is how dark photon events have been simulated in published FASER results so far~\cite{FASER:2023tle}.
\item \textit{With} spin correlations: The  dark photons that decay in the decay volume are boosted from the $A'$ rest frame to the $\pi^0$ rest frame. The $A'$s are then decayed to an $e^+ e^-$ pair, assigning the polar angle $\theta^*$ according to the distribution given in \cref{eq:angle-distribution-in-dark-photon-frame}.  The $e^-$ and $e^+$ are then boosted to the lab frame, and the resulting fermion energies in the lab frame have the distribution given in \cref{eq:pdensitywith}, as required once spin correlations are included. 
\end{itemize}

\begin{figure}[tb]
\centering
\includegraphics[width=\linewidth]{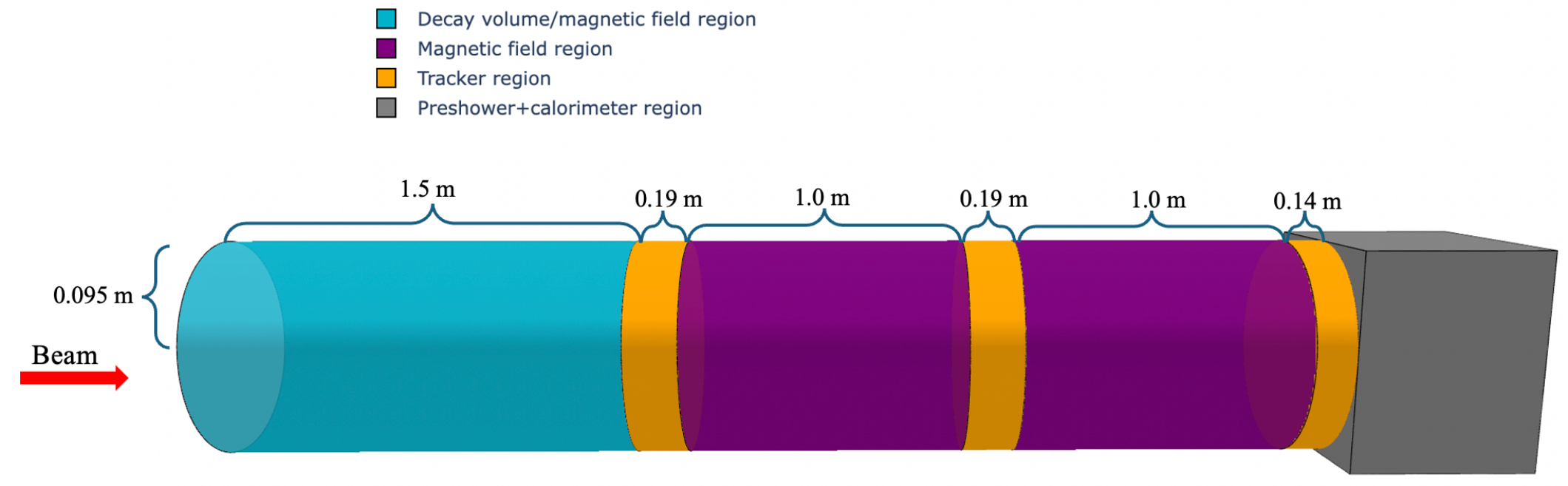}
\caption{The FASER detector geometry used in the ISG simulation (not to scale). Dark photons decay in the fiducial volume used by FASER~\cite{FASER:2023tle}, a cylindrical magnetic field region with radius 9.5 cm and length 1.5 m.  The decay products then pass through further magnetic field regions separated by tracker regions, as shown.  The preshower and calorimeter region is shown for reference, but is not included in the simulation. The detector is centered 6.5 cm above the beam collision axis to account for the beam half-crossing angle.} 
\label{fig:faser_det}
\end{figure}

The FASER detector has been described in detail in Ref.~\cite{FASER:2022hcn}. For this analysis, the most relevant parts of the detector geometry are summarized in \cref{fig:faser_det}.  Each of the dipole magnets creates a uniform magnetic field of $\bm{B} = (0.55~\text{T}, 0, 0)$ (pointing in the horizontal direction) in an empty cylindrical volume of radius 9.5 cm surrounding the axis of symmetry.  The inner core of the first magnet is considered to be the decay volume. Once a dark photon decays in this decay volume and the initial energies and momenta of the $e^-$ and $e^+$ are determined, we then propagate them downstream through the rest of the spectrometer until they enter the preshower or are deflected more than 9.5 cm from the axis of symmetry. We consider the effective detector to end at the last tracker region. Beyond the spectrometer, a detailed simulation of electromagnetic-shower development in the preshower station and its subsequent interaction in the calorimeter is required, but this is beyond the scope of this study. 
 
To isolate the $A' \to e^+ e^-$ signal and reduce background to a negligible level, a number of cuts were required in the FASER analysis~\cite{FASER:2023tle}.  For our purposes, the most relevant requirements are:
\begin{enumerate}
\setlength\itemsep{-0.05in}
\item {\em Total Energy.} The total energy deposited in the calorimeter is required to be $E_{\text{tot}} > 500~\gev$.
\item {\em Minimum Track Momentum.} There must be two charged tracks that each have momentum $p > 20~\gev$.
\item {\em Central Tracks.} Each of the two tracks must also be ``central,'' that is, remain within 9.5~cm of the detector's axis of symmetry until they enter the calorimeter.  
\end{enumerate}

\begin{figure}[tbp]
\includegraphics[width=0.98\textwidth]{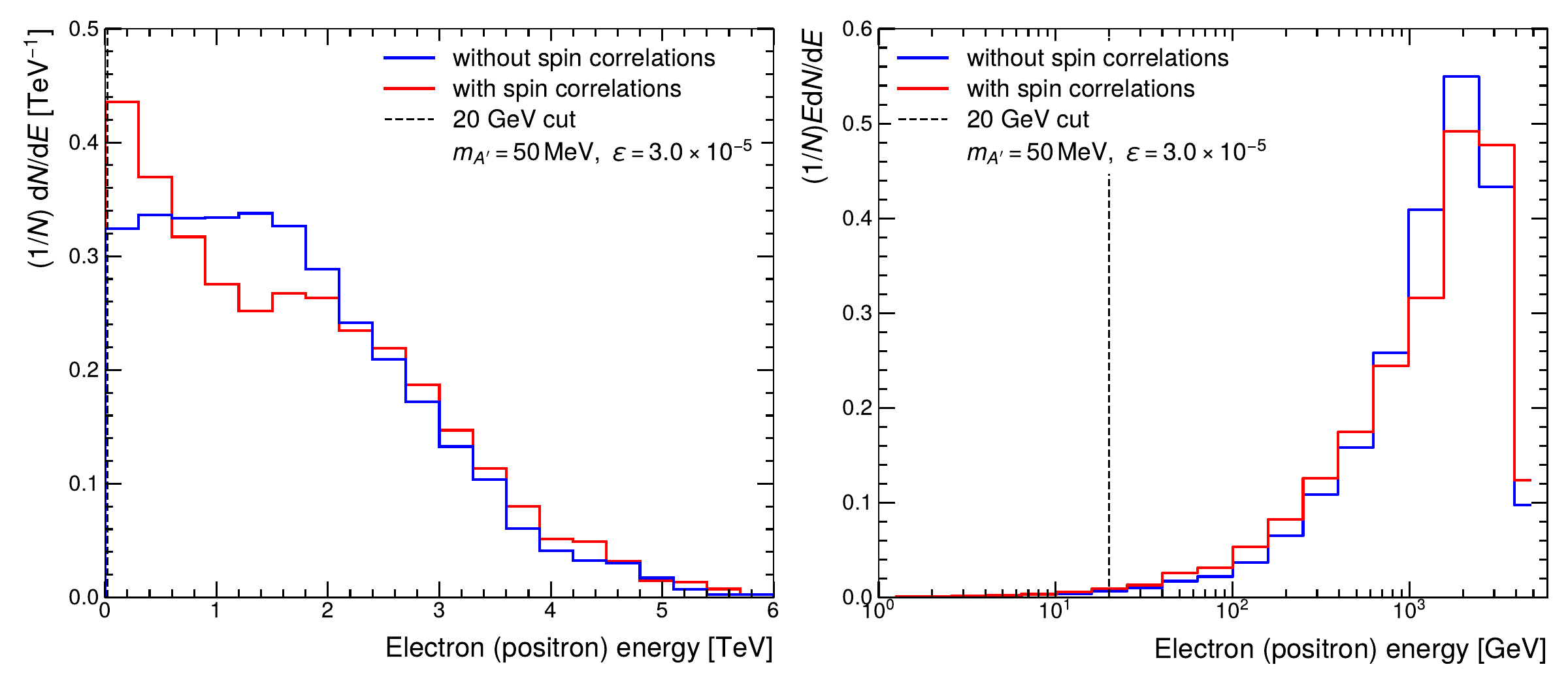}
\caption{Lab-frame energy spectra with (red) and without (blue) spin correlations for electrons that are produced by $\pi^0 \to \gamma A' \to \gamma e^+ e^-$ in the FASER decay volume.  The panels show the same simulation results, but with linear (left) and logarithmic (right) energy scales, and the dashed vertical line shows the 20 GeV minimum track momentum cut. The dark photon parameters are $(m_{A'}, \varepsilon) = (50~\text{MeV}, 3 \times 10^{-5})$. }
\label{fig:faser_rate}
\end{figure}

\begin{figure}[tbph]
\includegraphics[width=0.5\textwidth]{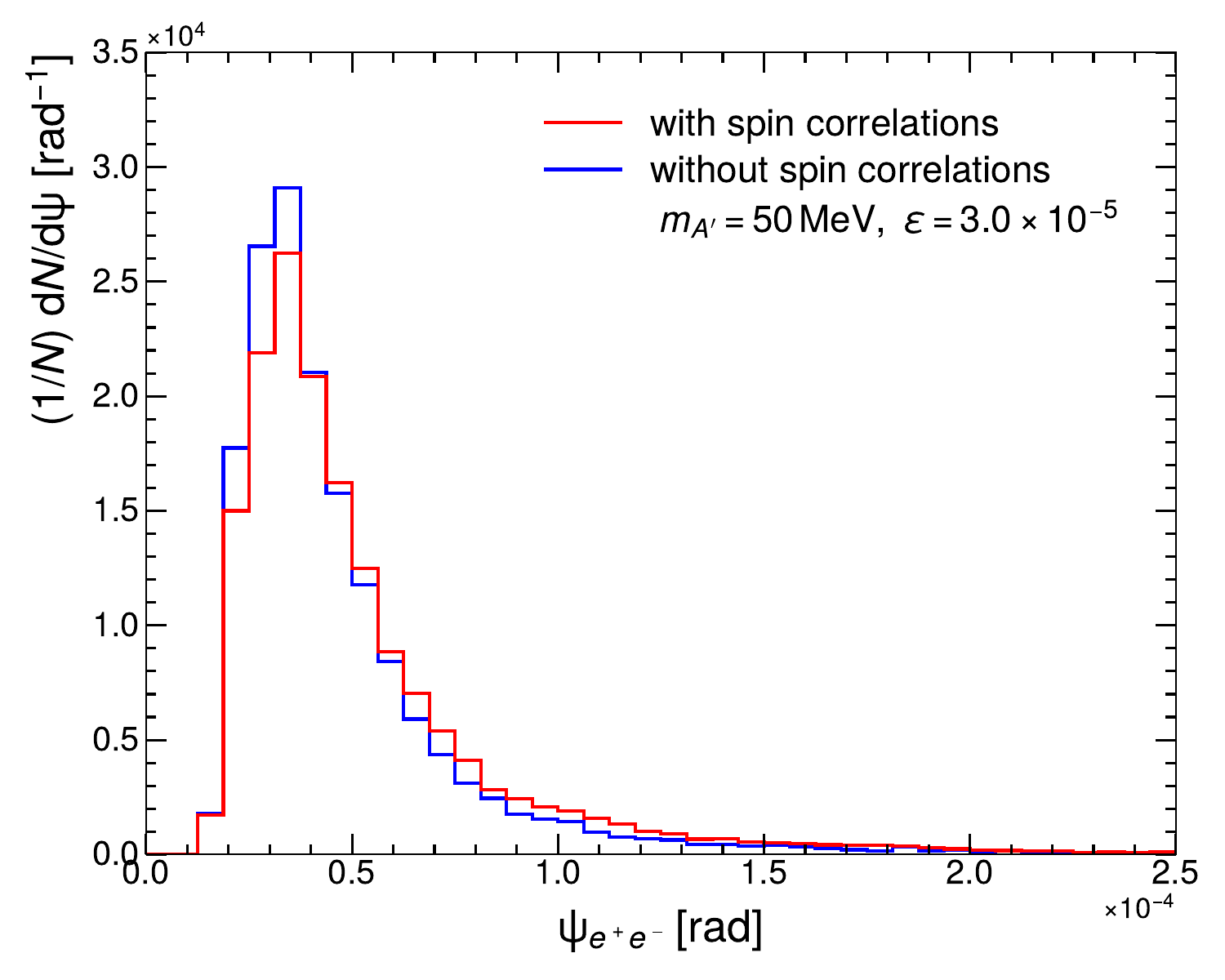}
\caption{The distribution of opening angles $\openingAngle_{e^+e^-}$ in the lab frame with (red) and without (blue) spin correlations for $e^+e^-$ pairs that are produced by $\pi^0 \to \gamma A' \to \gamma e^+ e^-$ and pass through the FASER decay volume.  The dark photon parameters are $(m_{A'}, \varepsilon) = (50~\text{MeV}, 3 \times 10^{-5})$.  On average, spin correlations slightly increase the opening angles.
}
\label{fig:faser_open_angle}
\end{figure}

We now consider the effect of spin correlations on event rates with these experimental cuts included.  We first consider the dark photon parameters $(m_{A'}, \varepsilon) = (50~\mev, 3 \times 10^{-5})$.  This point in parameter space is well-suited to this study, because for these parameters, dark photons are dominantly produced through $\pi^0$ decay and their only possible decay mode is to an $e^+ e^-$ pair. For this dark photon model, spin correlations have the following effects:
\begin{enumerate}
\setlength\itemsep{-0.05in}
\item {\em Total Energy.} Spin correlations do not alter the total energy of the $e^+ e^-$ pair, since this is fixed to be the energy of the decaying dark photon $E'$. Spin correlations therefore have a negligible effect on the number of events passing this cut.  
\item {\em Minimum Track Momentum.}  Spin correlations do affect the energies of the individual tracks, enhancing the energy asymmetry in the pion frame, as seen in \cref{fig:COM-pdfs}, and increasing the number of very soft tracks in the lab frame, as discussed in \cref{sec:decay-in-LAB-frame} and seen in \cref{fig:LAB-pdfs-pion}. For our representative dark photon model, \cref{fig:faser_rate} shows the $e^-$ energy spectra with and without spin correlations included. Spin correlations reduce the number of events passing this cut by $0.42\%$.  The effect is very small, because the number of events with soft tracks is a small fraction of all of the events.
\item {\em Central Tracks.} Spin correlations do not alter the polar angle distribution for a single decay product, but they do, on average, increase the opening angle between the decay products, as discussed in \cref{sec:angles-in-LAB-frame}.  This increases the probability that one of the fermions escapes the fiducial volume and so fails this requirement. \cref{fig:faser_open_angle} shows the opening angle distribution with and without spin correlations. As expected, spin correlations increase the average opening angle.  We find that spin correlations reduce the number of events that pass this central track cut by $0.88\%$. 
\end{enumerate}
There is naturally a correlation between events that fail cuts (2) and (3).  For our dark photon model with $(m_{A'}, \varepsilon) = (50~\mev, 3 \times 10^{-5})$, the overall effect of spin correlations is to reduce the number of signal events that pass all the cuts by approximately 1.0\%.  

\begin{figure}
\centering
\includegraphics[width=0.7\linewidth]{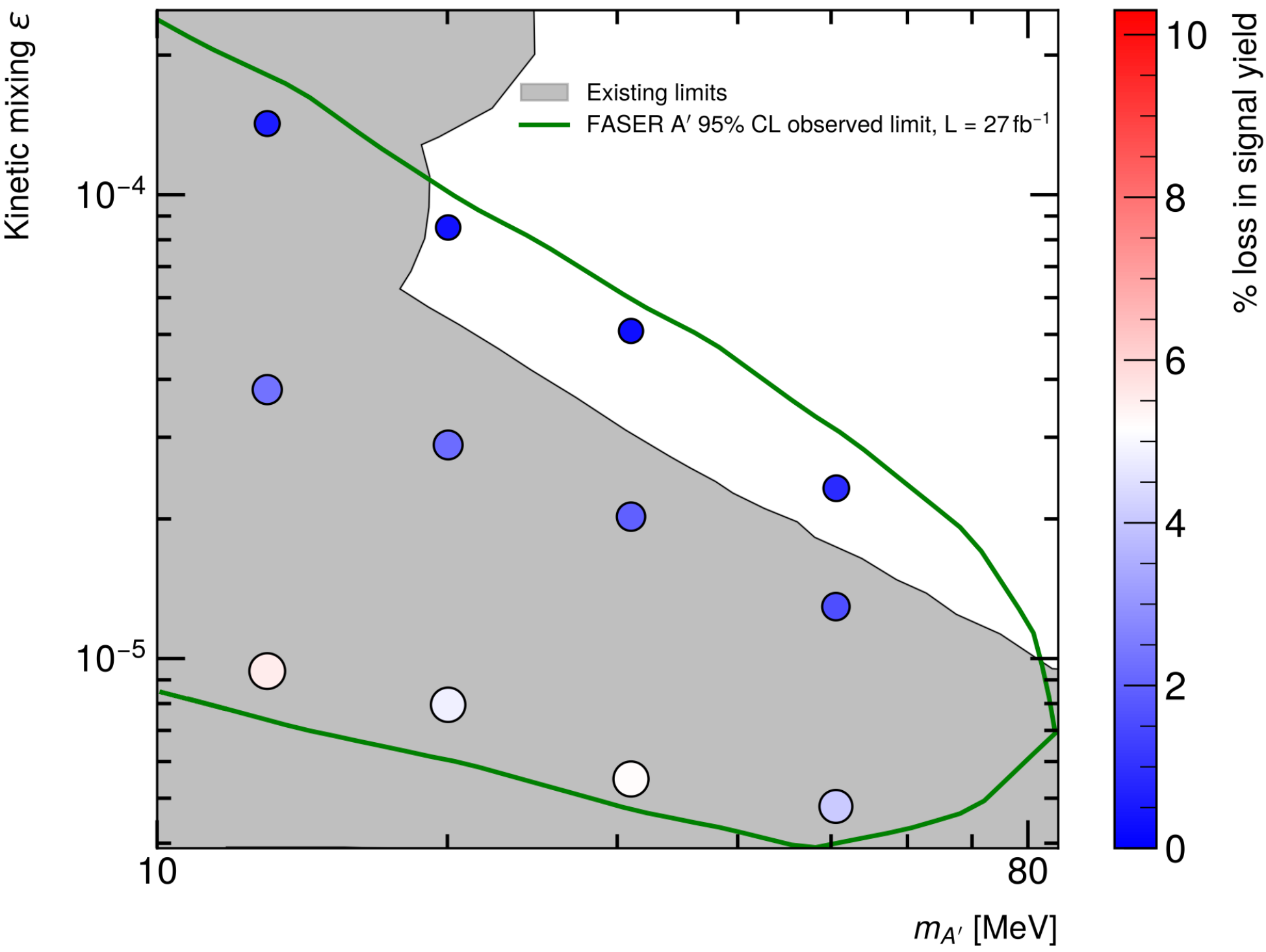}
\caption{The percent reduction in signal yield at FASER from including spin correlations for dark photons produced in $\pi^0$ decay for various points in the dark photon parameter space. FASER's observed limit on dark photons for 95\% CL with $L = 27~\text{fb}^{-1}$ (green) is shown. The gray shaded region is excluded by other existing limits.}
\label{fig:faser-SC-param-scan}
\end{figure}

In \cref{fig:faser-SC-param-scan}, we scan through the dark photon parameter space constrained by FASER. In general, the effect of spin correlations is greater on the Central Track cut than on the Minimum Track Momentum cut, although the effects are of a similar magnitude.  We see that throughout the parameter space with FASER sensitivity, spin correlations reduce the signal yield by 1\% to 6\%.  However, in the region where FASER has sensitivity beyond other current constraints, the reduction is only around 1\% to 2\%, a rather negligible reduction, and we expect that the effect will be negligible even for analyses based on the full Run 3 and HL-LHC integrated luminosities. Of course, if the experimental cuts are modified in future analyses, the effect of spin correlations may become more important and should be re-assessed.  FASER has also interpreted the signal yield in a search for axion-like particles as exclusion limits for dark photons~\cite{FASER:2024bbl}. We do not consider this result further, as it is not a dedicated dark photon analysis, but we note that, with no momentum selection applied, spin correlations can only affect geometric acceptance.

The future FASER2 experiment~\cite{Feng:2022inv} is currently planned to have a decay volume with a transverse aperture of $1~\text{m} \times 3~\text{m}$, and a longitudinal depth of approximately 10 meters. Because its solid angle coverage for dark photon decay is significantly larger than FASER's, the effect of spin correlations on angular cuts in FASER2 is expected to be much smaller than in FASER. From~\cref{fig:LAB-pdfs-pion,fig:faser-SC-param-scan} it can be inferred that increasing dark photon mass reduces the effect of spin correlations for a fixed momentum cut when the decay products are ultra-relativistic. Because there are no detailed dark photon analyses proposed for FASER2, we are unable to make a quantitative comparison at this time, but note that it would, of course, be important to consider the effect of spin correlations when detailed analyses are being planned. 

%%%%%%%%%%%%%%%%%%%%%%%%%%%%%%%%%
\subsection{Dark Photons at SHiP}
%%%%%%%%%%%%%%%%%%%%%%%%%%%%%%%%%

The SHiP experiment~\cite{Albanese:2878604} is a fixed target experiment planned for Experimental Cavern North 3 (ECN3) at CERN. The CERN SPS accelerator will deliver a high-intensity, 400 GeV proton beam to this location, with an expected $4 \times 10^{19}$ protons on target (PoT) per year, or $6 \times 10^{20}$ PoT for a 15-year run.  These fixed target collisions can produce light, feebly-interacting particles with larger rates, but smaller boosts, than can be achieved at the LHC. For dark photons, SHiP is projected to have sensitivity for masses $\sim 20~\text{MeV} - 4~\text{GeV}$ and couplings $\varepsilon \sim 1 \times 10^{-8} - 2 \times 10^{-4}$.  

Here we investigate the effect of spin correlations on the expected yield of dark photons produced by meson decays at SHiP. Meson decay is the dominant $A'$ production mechanism in SHiP's region of sensitivity up to $ m_{A'} \sim 500$~MeV. The majority of SHiP's sensitivity for dark photons in the currently-allowed parameter space with masses $m_{A'}<m_\eta$ comes from $\eta$ decay, and so we focus here on this production mode. 

To estimate the effect of spin correlations on the signal yield at SHiP, we simulate dark photon decays in SHiP with and without the effects of spin correlations using an adapted version of ISG. The two-dimensional momentum and angle distribution for $\eta$ production is extracted from Ref.~\cite{SHiP:2020noy}, which used SHiP's signal generator, FairShip~\cite{ShipSoft:FairShip2025}. The $\eta$ mesons are, then, decayed to dark photons, and the dark photons are decayed to $e^+ e^-$ and $\mu^+ \mu^-$ pairs.  To neglect spin correlations, the decays to fermions are isotropic in the $A'$ frame.  However, to include spin correlations, as described above for the FASER simulation, the $A'$ are boosted to the $\eta$ frame, the $A'$ are then decayed to fermions with the anisotropic angular distribution given in \cref{eq:angle-distribution-in-dark-photon-frame}, and the fermions are then boosted back to the lab frame.  

\begin{figure}[tbp]
    \centering
\includegraphics[width=1.0\linewidth]{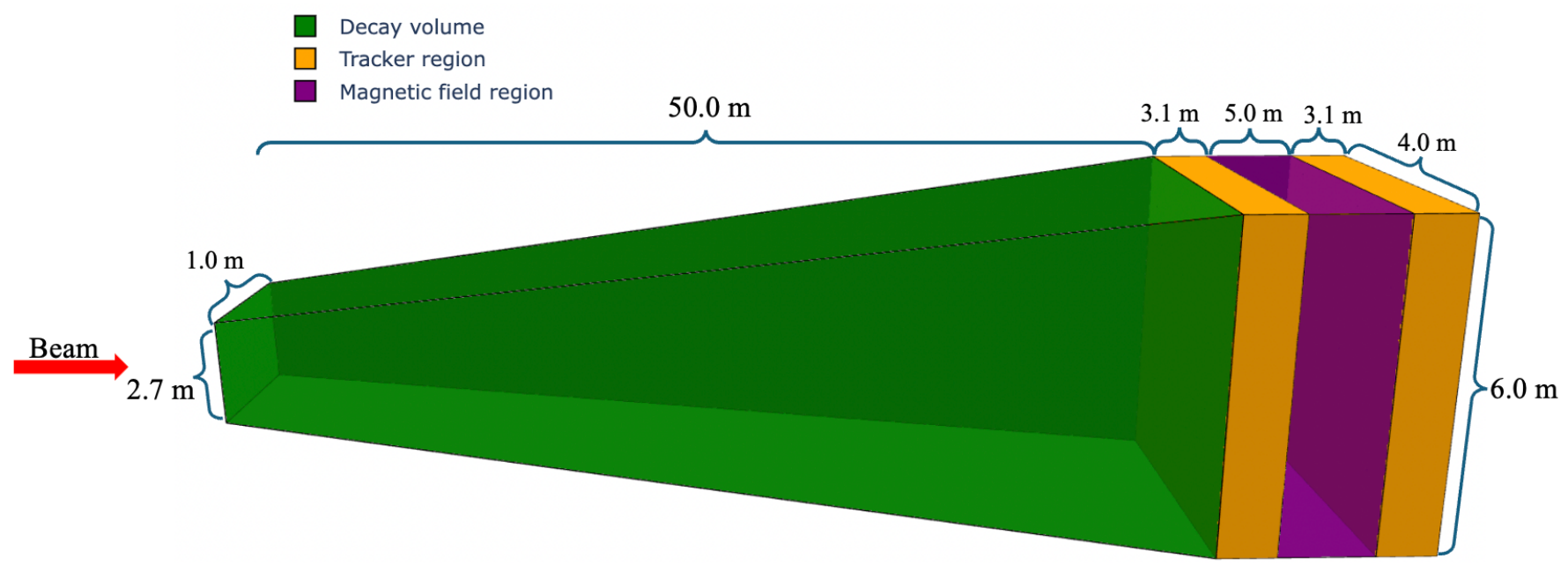}
\caption{The SHiP detector geometry used in the ISG simulation (not to scale). The detector has a fiducial decay volume in the shape of a pyramidal frustum with a length of 50 m, upstream dimensions of $1.0~\text{m} \times 2.7~\text{m}$, and downstream dimensions of $4.0~\text{m} \times 6.0~\text{m}$. Downstream of the decay volume there is a 3.1 m-long tracker region, a 5.0 m-long magnetic field region, and another 3.1 m-long tracker region ending at the last tracking station, which we consider the end of the detector. Each tracker region represents two tracking stations.  The beam is aligned with the detector axis of symmetry. }
    \label{fig:ship_det}
\end{figure}

The SHiP detector is described in detail in Ref.~\cite{Albanese:2878604}.  For this analysis, the most relevant parts of the detector geometry are summarized in \cref{fig:ship_det}. Dark photons are forced to decay within SHiP’s decay volume, with decay positions distributed exponentially according to the $A'$ decay length.  We then propagate each decay product until it either exits the side of the detector or reaches the end of the last tracker region. We assume a constant magnetic field $\bm{B} = (0.14~\text{T},0,0)$ in the magnetic field region. 

Using this simulation of SHiP in ECN3, we can estimate the impact of spin correlations on the efficiency of the SHiP dark photon search.  SHiP employs a number of cuts to isolate the $A' \to e^+ e^-, \mu^+ \mu^-$ signal and reject background events; see Ref.~\cite{Albanese:2878604}. For this analysis, the most relevant requirements are: 
\begin{enumerate}
\setlength\itemsep{-0.05in}
\item {\em Impact Parameter of $A'$ Decay.} The pair of tracks must extrapolate back to a vertex that is within 10 cm of the fixed target collision axis and more than 5 cm from the boundary of the decay volume.
\item {\em Minimum Track Momentum.}  There must be two charged tracks that each have momentum $p >1.0~\text{GeV}$.
\item {\em Central Tracks.}  The tracks must remain in the $4.0~\text{m} \times 6.0~\text{m}$ physics aperture until they exit the last tracker region. 
\end{enumerate}

We can now examine the effect of spin correlations on the efficiencies of each of these cuts.  For SHiP, we first consider the dark photon parameters $(m_{A'},\varepsilon) = (300~\text{MeV}, 5 \times 10^{-7})$.  This point uses a benchmark dark photon mass used by the SHiP Collaboration and is at a log-central point in $\varepsilon$ for SHiP's reach. For this point, the dark photon has a roughly equal chance of decaying to $e^+e^-$ and $\mu^+\mu^-$, with $B(A'\to e^+e^-)=53\%$ and $B(A'\to \mu^+\mu^-)=47\%$.

\begin{figure}[tbp]
\centering
\includegraphics[width=0.98\linewidth]{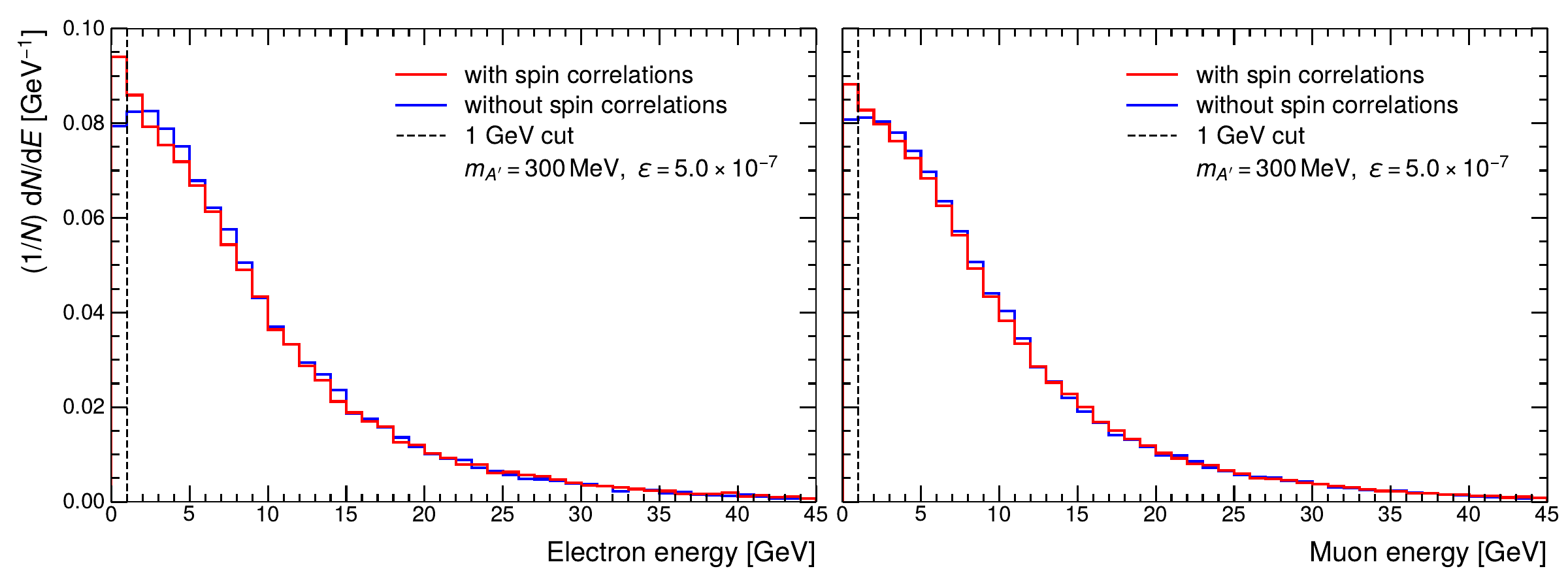}
\caption{Lab-frame energy spectra with (red) and without (blue) spin correlations for electrons (left) and muons (right) that are produced by $\eta \to \gamma A' \to \gamma e^+ e^-, \gamma \mu^+ \mu^-$ in the SHiP decay volume, within 10 cm of the beam dump collision axis, and more than 5 cm from the boundaries of the decay volume, as required by the impact parameter cut.  The dashed vertical line shows the 1.0 GeV minimum track momentum cut.  The dark photon parameters are $(m_{A'}, \varepsilon) = (300~\text{MeV}, 5 \times 10^{-7})$.}
\label{fig:ship_lepton_energy}
\end{figure}

\begin{figure}[tbp]
\centering
\includegraphics[width=0.98\linewidth]{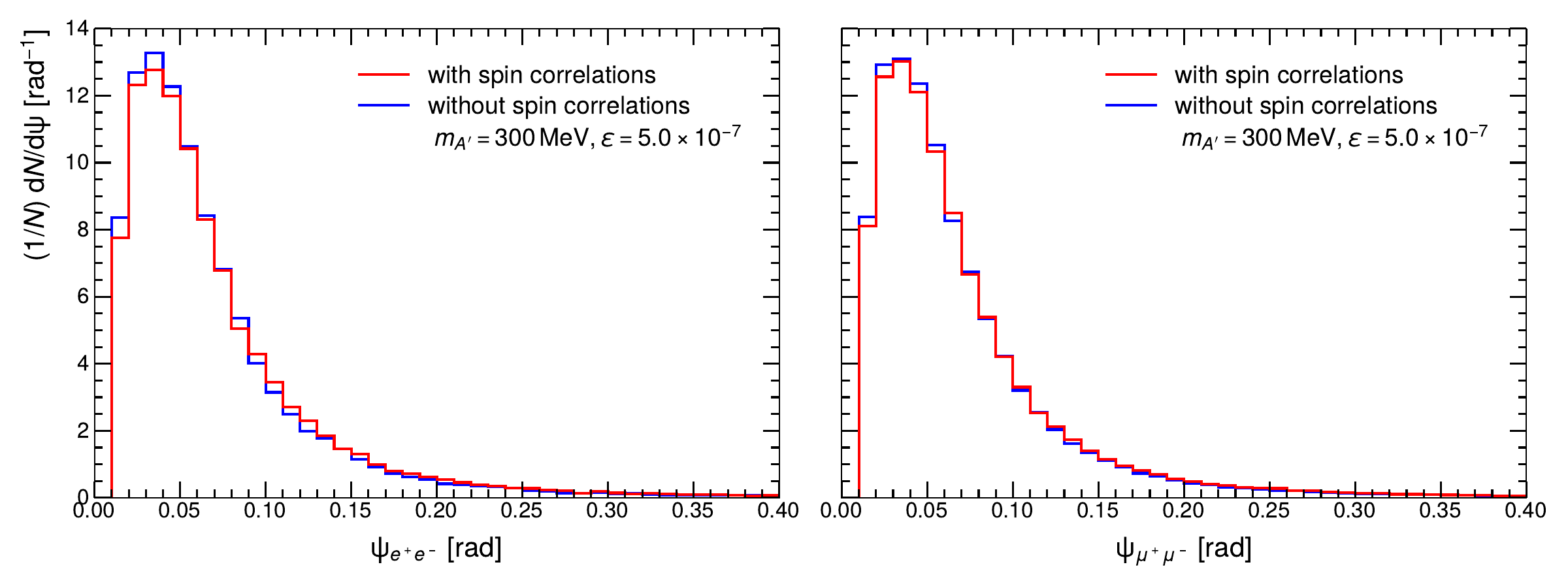}
\caption{The distribution of opening angles $\openingAngle_{e^+e^-}$ (left) and $\openingAngle_{\mu^+ \mu^-}$ (right) in the lab frame with (red) and without (blue) spin correlations for fermion pairs that are produced by $\eta \to \gamma A' \to \gamma e^+ e^-, \gamma \mu^+ \mu^-$ in the SHiP decay volume, within 10 cm of the beam dump collision axis, and more than 5 cm from the boundaries of the decay volume, as required by the impact parameter cut. The dark photon parameters are $(m_{A'}, \varepsilon) = (300~\text{MeV}, 5 \times 10^{-7})$. }
\label{fig:ship_open_angle}
\end{figure}

\begin{enumerate}
\setlength\itemsep{-0.05in}
\item {\em Impact Parameter of $A'$ Decay.} Spin correlations have no effect on the location of $A'$ decays.
\item {\em Minimum Track Momentum.}  The fermion energy spectra for $e^-$ and $\mu^-$ with and without spin correlations are shown in \cref{fig:ship_lepton_energy}.  As with FASER, spin correlations soften these energy spectra, and so reduce the number of events that pass the minimum track momentum cut.  In contrast to FASER, however, since the typical dark photon energy at SHiP is $\mathcal{O}(10~\text{GeV})$, the 1 GeV momentum cut on each decay product represents a substantial fraction. For the chosen signal model, we find that the cut on momentum results in a 7.7\% reduction in electron events and a 2.4\% reduction in muon events. 
\item {\em Central Tracks.}  The opening angle distributions with and without spin correlations are shown in \cref{fig:ship_open_angle}.  Spin correlations slightly increase the typical opening angles, as anticipated in \cref{sec:angles-in-LAB-frame}, and so can increase the number of events where the fermions leave the detector before the spectrometer.  The SHiP detector has a larger transverse area than FASER, but it also has far less boosted decay products, so we anticipate a non-negligible loss in yield due to the decay products leaving out of the side of the detector. We find that spin correlations result in a 8.7\% reduction in electron events and a 1.9\% reduction in muon events. 
\end{enumerate}

Accounting for both the momentum cut and loss of events due to exits from the detector, as well as the high correlation between these cuts, there is a total reduction of 9.8\% for electron events and 2.7\% for muon events. As expected given the analytical results above, leading to \cref{eq:discrepancy}, the effect of spin correlations is maximal for light fermions, and so is larger for electrons than for muons.  Given the branching fractions for this dark photon model, spin correlations produce a total 6.5\% reduction in signal yield from $\eta$ decay, the dominant production mode at this point in parameter space. 

\begin{figure}[tbp]
\centering
\includegraphics[width=0.8\linewidth]{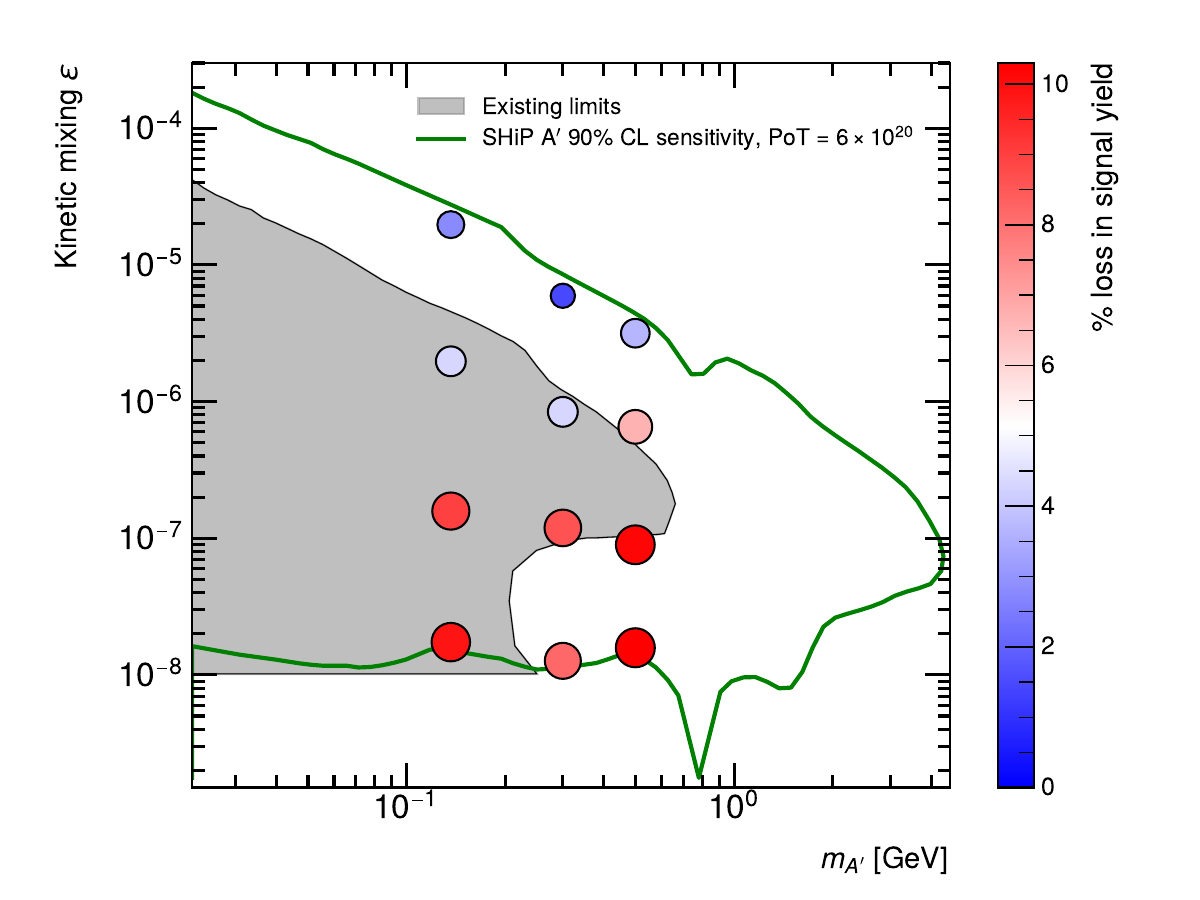}
\caption{
The percent reduction in signal yield at SHiP from including spin correlations for dark photons produced in $\eta$ decay for various points in the dark photon parameter space. The green contour is SHiP's 90\% CL sensitivity limit to dark photons, neglecting spin correlations, assuming $6\times10^{20}$ PoT, and including all production modes.  The gray region is excluded by existing limits.}
\label{fig:ship_param_scan}
\end{figure}

We now scan through the dark photon parameter space where SHiP has sensitivity; the results are shown in \cref{fig:ship_param_scan}.  We find that once spin correlations are included, the loss in events due to fermions leaving through the side of the detector is always comparable to the loss from the cut on the momentum, and often removes the same events.  As seen in \cref{fig:ship_param_scan}, spin correlations reduce the signal efficiency for dark photons produced in $\eta$ decays by 1\% to 10\% in SHiP's region of sensitivity.  The largest effect is in the region with low coupling $\varepsilon$, where a $\sim 10\%$ loss in signal efficiency is significant.  The impact of spin correlations may, of course, be enhanced or suppressed by alternative experimental cuts, and this impact would be worth re-visiting in the future as these cuts become better defined. 

%****************************************
\section{Conclusions} 
\label{sec:conclusions}
%****************************************

Long-lived particles are important targets of BSM searches, and dark photons are among the most prominent examples.  In many experimental settings, dark photons with masses between 10 MeV and 1 GeV are dominantly produced in pseudoscalar meson decays, for example, $\pi^0 \to \gamma A'$.  They then travel long distances before decaying, for example, through $A' \to e^+e^-$. Because dark photons are long-lived, the narrow-width approximation can be applied to the dark photon propagator. However, it is not true that dark photon production and decay can be treated separately, as there are spin correlations between production and decay. In this work, we have analyzed the impact of these spin correlations on the energy and angular distributions of dark photon decay products, deriving analytic formulas to describe these distributions and incorporating spin correlations into event generators to assess their impact on signal event rates.

The origin of spin correlations in these processes is easy to understand.  Dark photons are massive vector bosons, and so they have two transverse polarizations and one longitudinal polarization. When spin correlations are ignored, dark photons are assumed to be produced unpolarized.  However, dark photons produced in pseudoscalar meson decays are never produced with a longitudinal polarization, because photons only have transverse polarizations and pions are scalars. 

In this work, we have shown that the absence of the dark photon's longitudinal polarization increases the probability that the $e^-$ and $e^+$ are produced parallel and anti-parallel to the dark photon's momentum in the meson frame.  The total energy of the $e^+ e^-$ system is unaffected, but spin correlations increase the asymmetry of the $e^-$ and $e^+$ energies.  More precisely, spin correlations modify the electron energy spectrum in the pion frame from the flat distribution of \cref{eq:pdensitywithout} to the convex parabola of \cref{eq:pdensitywith}, increasing the number of very soft and very hard tracks. 

To quantify the effect of spin correlations on the electron energy spectrum, we defined a discrepancy parameter $\Delta$ in \cref{eq:DeltaSpecific}, which is Lorentz invariant and has the wonderfully simple form given in \cref{eq:discrepancy}. It is maximized for light decay products, like electrons, and its maximal value $\frac{1}{6\sqrt{3}}\approx 9.6\%$ provides a rough estimate of the size of spin correlation effects. We used the probability density calculus developed in \cref{appendix:pdf-calculus-intro} to boost the electron spectrum from the pion frame into the lab frame, resulting in the analytic expressions given in \cref{sec:decay-in-LAB-frame} and the electron energy distributions shown in \cref{fig:LAB-pdfs-pion}. 

In \cref{sec:angulardistributions} we explored the impact of spin correlations on angular distributions.  Using probability density calculus, we computed the fermion opening angle distribution in the meson frame in the massless fermion limit (\cref{appendix:specific-angular-pdf}), as well as the fermion opening angle distribution in the lab frame in the ultra-relativistic limit (\cref{eq:LAB-angular-pdf}). Using these analytic techniques, we found that, although spin correlations do not alter the polar angle distribution for a single decay product, they do, on average, increase the opening angle between the decay products, as shown in \cref{eq:opening-angle-enhancement}, which also has experimental consequences. 

To investigate the effects of spin correlations in more realistic experimental analyses, we have shown how to include spin correlations in event generators using a semi-classical approach that only modifies the angular distribution of the decay products in dark photon decays (see \cref{eq:angle-distribution-in-dark-photon-frame}), but still reproduces the correct analytic result. This approach provides a simple way to include spin correlations in existing signal simulators, like FORESEE~\cite{Kling:2021fwx} and FairShip~\cite{ShipSoft:FairShip2025}. We then used this approach to estimate the effect of spin correlations on published limits on dark photons from FASER, as well as projected searches at SHiP.  We found that spin correlations have a negligible effect on FASER's published limits, but may decrease SHiP's signal yield by $\sim 10\%$ in parts of dark photon parameter space.  In the case of FASER, since collider energies are large, the cut on final-state particle momenta at 20 GeV is far below the energies where the asymmetry is maximal around 600 GeV, as seen in \cref{fig:faser_rate}.  In contrast, at SHiP, given the lower fixed-target energies, the cut on final state particle momenta at 1 GeV is close to 2 GeV, where the spin correlation effect is maximal, as seen in see \cref{fig:ship_lepton_energy}.
Spin correlations, then, may be significant, depending on the experimental cuts implemented, and they should be included in future analyses.

Last, we note that our analysis has been confined to spin correlations in searches for dark photons produced in meson decays.  We have not analyzed the spin correlations from other production mechanisms, such as dark bremsstrahlung.  We note also that these spin correlations are also present in searches for other long-lived particles, such as other light gauge bosons and heavy neutral leptons. From the point of view of BSM searches, large spin correlation effects may be seen as a nuisance, as they complicate the analysis.  However, in the event of a discovery, the spin correlation effects studied here may provide a welcome avenue to help determine the spin and other characteristics of the newly-discovered particle.

%****************************************
\section*{Acknowledgments}
%****************************************

We thank Tao Han and Felix Kling for insightful conversations.  This work is supported in part by U.S.~National Science Foundation Grants PHY-2111427, PHY-2210283, and PHY-2514888 and Simons Foundation Grant 623683.  The work of J.L.F.~is supported in part by Simons Investigator Award 376204, and Heising-Simons Foundation Grants 2019-1179 and 2020-1840.  

\appendix

%****************************************
\section{Probability Density Calculus}
\label{appendix:pdf-calculus-intro}
%****************************************

In this Appendix, we summarize the essentials of probability density calculus used to analyze particle decays in this paper. Following convention, we denote random variables by capital letters, for example, $X, Y, Z$. Specific values of a random outcome are denoted by lowercase letters, $x, y, z$. We say that a random variable $X$ is distributed according to a probability density function $\pdensity_X$. This means that the probability of the random variable $X$ attaining a value between $x$ and $x + \de x$ is $\pdensity_X(x)\de x$. 

Suppose there is a new random variable, $Y = f(X)$, which is a function of the previous random variable $X$. Then, the probability density of $Y$ is
\begin{align}\label{appendix:dependent-pdf-formula}
    \pdensity_Y(y) & = \int\pdensity_X(x) \, \delta(y - f(x)) \, \de x\,.
\end{align}
The proof of this is as follows. For a moment let's assume $f$ is monotonic. If we generate $N$ events of variable $X$, then $N' = N \, \pdensity_X(x) \, |\de x|$ of them will fall between the values $x$ and $x + \de x$. Once we map these $N'$ values with the function $f$, we will get the same number of events in the interval from $y$ to $y + \de y$ because $f$ is monotonic, and so $N' = N\pdensity_Y(y)|\de y|$, where $y = f(x)$. From this it follows that
\begin{align}\label{appendix:dependent-pdf-formula-little}
    N \, \pdensity_X(x) \, |\de x| & = N \, \pdensity_Y(y) \, |\de y|\,,
\end{align}
and so
\begin{align}
    \pdensity_Y(y) & = \frac{\pdensity_X\left(f^{-1}(y)\right)}{\left|f'(f^{-1}(y))\right|}\,.
\end{align}
In the case of non-monotonic functions $f$, we must sum over all $X$ values, $x_0$, which get mapped to the same value $y = f(x_0)$.
\begin{align}
    \pdensity_Y(y) & = \sum_{x_0\,:\,f(x_0)\,=\,y} \frac{\pdensity_X\left(x_0\right)}{\left|f'(x_0)\right|} = \int\pdensity_X(x) \, \delta(y - f(x)) \, \de x\,.
\end{align}
In the last step we used the Dirac delta function identity $\delta\left(y - f(x)\right) = \sum_{x_0\,:\,f(x_0)\,=\,y}\frac{\delta\left(x - x_0\right)}{\left|f'(x_0)\right|}\,.$\\
If $Y$ is a function of multiple random variables $Y = f(X_1, \dots, X_n)$, we can generalize \cref{appendix:dependent-pdf-formula} as
\begin{align}
\label{appendix:dependent-pdf-formula-big}
    \pdensity_Y(y) & = \int\pdensity_{X_1, \dots, X_n}(x_1,\dots, x_n) \, \delta(y - f(x_1, \dots, x_n)) \, \de x_1\dots\de x_n\,.
\end{align}
Through this paper energy distributions are denoted by the letter $\pdensityEnergy$ without any further descriptive subscript. Angular distributions are denoted with $\pdensity_\Theta$ and $\pdensity_\Phi\,.$

%****************************************
\section{Boosting Particle Energy Spectra}
\label{appendix:pdf-calculus-1to2-decay}
%****************************************

In this Appendix, we will illustrate the probability density calculus on the example of a dark photon decaying into two fermions, first in the dark photon frame and then in the lab frame. The result of this exercise will be a general recipe how to boost any energy spectrum into a different frame, culminating in \cref{appendix:general-boosted-pdf}.

Particle masses and momenta in the lab frame will be labeled as in \cref{fig:pion-decay}. The corresponding quantities in the dark photon frame will be labeled with asterisks, e.g., $E_i^*$ and $k_i^*$. In the dark photon frame, both fermions are mono-energetic, and their energy distributions are 
\begin{align}\label{eq:sharp}
    \pdensity^*\left(E_i^*\right) & = \delta\left(E_i^* - \frac{m}{2}\right)\,.
\end{align}

To determine the fermion energy probability distribution in the lab frame, we boost the fermion momenta in the dark photon frame by $\gamma = E_{\bm{q}}/m$ in the direction of $\bm{q}$. $E_{\bm{q}}$ is the dark photon's energy in the lab frame, now not the same as in \cref{eq:energies}. We can take $\bm{q}$ to be along the $z$ axis. Fermion momenta in the dark photon frame are determined by two uniformly sampled spherical angles $\Theta^*$ and $\Phi^*$. The electron energy in the lab frame
\begin{align}
    E_1 & = \gamma\left(E_1^* + \beta k_1^*\cos\Theta^*\right) = \gamma\left(E_1^* + \beta\cos\Theta^*\sqrt{(E_1^*)^2 - m_f^2}\ \right)\,,
\end{align}
is now a random variable which is a function of three other random variables, $E_1 = E_1\left(E_1^*, \Theta^*, \Phi^*\right)\,.$ The spherical angles have uniform probability distributions $\pdensity_{\Phi}^*(\phi^*) = \frac{1}{2\pi} 
\quad \textrm{and} \quad
\pdensity_\Theta^*(\theta^*) = \frac{1}{2}\sin\theta^*\,.$
Applying \cref{appendix:dependent-pdf-formula-big}, the probability density for the electron energy in the lab frame is 
\begin{align}
    \pdensity(E_1) & = \int\pdensity_{\Phi}^*(\phi^*)\pdensity_\Theta^*(\theta^*)\pdensity^*(E_1^*)\delta\left(E_1 - \gamma\left(E_1^* + \beta\cos\theta^*\sqrt{(E_1^*)^2 - m_f^2}\right)\right)\de E_1^*\de\theta^*\de\phi^*\nonumber\\
    & = \int\frac{1}{2}\sin\theta^* \ \pdensity^*(E_1^*) \ \delta\left(E_1 - \gamma\left(E_1^* + \beta\cos\theta^*\sqrt{(E_1^*)^2 - m_f^2}\right)\right)\de E_1^*\de\theta^*\,.\label{appendix:even-more-general-boosted-pdf-before-simplifying}
\end{align}
Integrating with respect to $\theta^*$, we find 
\begin{align}\label{appendix:general-boosted-pdf}
    \pdensityEnergy(E_1) & = \frac{1}{2\gamma\beta}\int_{E_1^* = \gamma\left(E_1 - \beta\sqrt{E_1^2 - m_f^2}\right)}^{E_1^* = \gamma\left(E_1 + \beta\sqrt{E_1^2 - m_f^2}\right)}\frac{\pdensity^*(E_1^*)\de E_1^*}{\sqrt{(E_1^*)^2 - m_f^2}}\,.
\end{align}
The only assumption in the derivation of this equation was that $\pdensity^*(E_1^*)$ has no angular dependence. Therefore, equation \cref{appendix:general-boosted-pdf} is quite general and can be used to boost energy spectra from one frame to another. In the case where the energy spectrum is anisotropic, \cref{appendix:even-more-general-boosted-pdf-before-simplifying} must be used together with $\rho^* = \rho^*\left(E_1^*, \cos\Theta^*\right)$. Substituting \cref{eq:sharp} into \cref{appendix:general-boosted-pdf}, we get
\begin{align}
    \pdensityEnergy(E_1) & = \frac{1}{2\gamma\beta\sqrt{\frac{m^2}{4} - m_f^2}}\Theta \! \left( \! E_1 \! - \! \gamma \! \left(\frac{m}{2} - \beta\sqrt{\frac{m^2}{4} - m_f^2}\right) \! \! \right) \Theta \! \left( \! \gamma \! \left(\frac{m}{2} + \beta\sqrt{\frac{m^2}{4} - m_f^2}\right) \! - \! E_1 \! \right) \! . %%\gamma\left(\frac{m}{2} - \beta\sqrt{\frac{m^2}{4} - m_f^2}\right) < \varepsilon < \gamma\left(\frac{m}{2} + \beta\sqrt{\frac{m^2}{4} - m_f^2}\right)\, .
    \label{eq:temp-4}
\end{align}
We see that a sharp, delta function energy distribution has been boosted into a flat, uniform plateau of energies with sharp cut-offs. We will use this intuitive description to understand why plateaus generally appear in energy spectra. In \cref{appendix:why-is-there-a-plateau?}, we show that by chopping up the initial energy spectrum into approximate delta functions, boosting each delta function individually, and then summing them back up, we can understand how plateaus arise in more involved decays.

%****************************************
\section{Particle Decay Example: \texorpdfstring{\bm{${\pi^0\to\gamma A'\to\gamma e^+e^-}$}}{}}
\label{appendix:pdf-calculus-three-body}
%****************************************

In this section we will show that a purely classical analysis of pion decay in the pion frame yields a uniform electron energy spectrum, and that this precisely describes the situation where spin correlations are neglected, \cref{eq:pdensitywithout}.

In the pion frame, the photon and dark photon energies have fixed values shown in \cref{eq:energies}. In the dark photon frame, the electron and positron have fixed energies and uniformly distributed momentum directions.

To determine the electron energy distribution in the pion frame, we must boost the electron energy distribution from the dark photon frame into the pion frame by $\gamma = \frac{E_{\bm{q}}}{m} = \frac{E'}{m}$ in the direction of the dark photon momentum in the pion frame, which we denote $\bm{q}$. The dark photon momentum in the pion frame is an isotropic random variable.

We will again adopt the labeling conventions of \cref{fig:pion-decay}, but now these labels denote quantities in the pion frame, and we denote quantities in the dark photon frame with asterisks, so, for example, $k_1^{*\mu}$ is the electron 4-momentum in the dark photon frame.  We let $q^\mu = \left(E', \bm{q}\right), k_1^{*\mu} = \left(E_1^*, \bm{k}_1^*\right), k_1^\mu = \left(E_1, \bm{k}_1\right), E_1^* = m/2\,.$ Crucially, momenta $\bm{q}$ and $\bm{k}_1^*$ are distributed uniformly on a sphere. Let $\bm{n}\left(\Theta, \Phi\right)$ be the unit vector parametrized by spherical angles $\Theta$ and $\Phi$. Then
\begin{align}
    \bm{q} & = q\bm{n}\left(\Theta, \Phi\right) \ , \quad \bm{k}_1^* = k_1^*\bm{n}\left(\Theta^*, \Phi^*\right) \ , \quad q = E' \ , \quad k_1^*= \sqrt{\frac{m^2}{4} - m_f^2} \ . \label{eq:temp-6}
\end{align}
We boost $k_1^{*\mu}$ by $\gamma = \frac{E'}{m} = \frac{M^2 + m^2}{2mM}$ in the direction of $\bm{q}$ to get $E_1$, 
\begin{align}
    E_1 & = \gamma\left(E_1^* - \frac{\bm{q}\cdot\bm{k}_1^*}{\gamma m}\right) = \gamma\left(E_1^* - \frac{qk_1^*}{\gamma m}\bm{n}\left(\Theta, \Phi\right)\cdot\bm{n}\left(\Theta^*, \Phi^*\right)\right)\,.
\end{align}
The probability density describing the electron energy in the pion frame is, using \cref{appendix:dependent-pdf-formula},
\begin{align}\label{temp-5}
    \pdensityEnergy(E_1) = \int\frac{\sin\theta\de\theta}{2}\frac{\sin\theta^*\de\theta^*}{2}\frac{\de\phi}{2\pi}\frac{\de\phi^*}{2\pi}\delta\left(E_1 - \gamma\left(E_1^* - \frac{qk_1^*}{\gamma m}\bm{n}\left(\theta, \phi\right)\cdot\bm{n}\left(\theta^*, \phi^*\right)\right)\right)\,.
\end{align}
This integral can be calculated piece by piece using spherical symmetry. Writing
\begin{align}\label{temp-6}
    \pdensityEnergy(E_1) = \int\frac{\sin\theta\de\theta}{2}\frac{\de\phi}{2\pi} I(\theta, \phi, E_1)\,,
\end{align}
we see that $I$ is independent of direction, and so
\begin{align}
    I(\theta, \phi, E_1) & = \int\frac{\sin\theta^*\de\theta^*}{2}\frac{\de\phi^*}{2\pi}\delta\left(E_1 - \gamma\left(E_1^* - \frac{qk_1^*}{\gamma m}\bm{n}\left(\theta, \phi\right)\cdot\bm{n}\left(\theta^*, \phi^*\right)\right)\right) = I(0, 0, E_1)\nonumber\\
    & = \int\frac{\sin\theta^*\de\theta^*}{2}\frac{\de\phi^*}{2\pi}\delta\left(E_1 - \gamma\left(E_1^* - \frac{qk_1^*}{\gamma m}\cos\theta^*\right)\right)\nonumber\\
    & = \frac{1}{2}\int_{-1}^{1}\delta\left(E_1 - \gamma\left(E_1^* - \frac{q k_1^*}{\gamma m}u\right)\right)\de u\,.
\end{align}
We do not need to compute this integral precisely, but just note that it is a constant throughout some allowed range of $E_1$. That constant must be the reciprocal value of the length of the allowed energy interval, $E_{1\textrm{min}} < E_1 < E_{1\textrm{max}}$. The maximum and minimum values of $E_1$ can be read off from the integrand and are given by
\begin{align}
    \gamma\left(E_1^* \pm \frac{qk_1^*}{\gamma m}\right) & = E_{1\textrm{max}/1\textrm{min}} = \frac{1}{2}\left(E' \pm E\sqrt{1 - \frac{4m_f^2}{m^2}} \ \right)\,.
\end{align}
This recovers the exact result of \cref{eq:pdensitywithout}
\begin{align}
    \pdensityEnergy(E_1) & = \cfrac{1}{E_{1\textrm{max}} - E_{1\textrm{min}}} \ \Theta\left(E_1 - E_{1\textrm{min}}\right) \ \Theta\left(E_{1\textrm{max}} - E_1\right)\nonumber\\
    & = \frac{2M}{M^2 - m^2}\frac{1}{\sqrt{1 - \frac{4m_f^2}{m^2}}} \ \Theta\left(E_1 - E_{1\textrm{min}}\right) \ \Theta\left(E_{1\textrm{max}} - E_1\right)\,. 
\end{align}
We have thus shown that neglecting spin correlations amounts to a purely classical analysis of the decay.

%****************************************
\section{Electron Energy Spectrum Features in the Lab Frame}
\label{appendix:why-is-there-a-plateau?}
%****************************************

In this Appendix, we provide a qualitative description of the electron energy spectrum in the lab frame given by \cref{fig:LAB-pdfs-pion}.  The application of \cref{appendix:general-boosted-pdf} results in \cref{eq:probability-density-function-without-spin-correlations-lab-frame,eq:probability-density-function-with-spin-correlations-lab-frame}, but how can we physically understand this result?

In \cref{appendix:pdf-calculus-1to2-decay} we showed that boosting a sharply peaked delta function electron energy distribution associated with a $1\to 2$ decay gives a uniform rectangular distribution. For the sake of simplicity, let us analyze the spectrum without spin correlations. The results in \cref{fig:LAB-pdfs-pion} can be intuitively understood as follows. In the pion frame, the electron energy distribution looks like a uniform distribution with sharp cutoffs at $E_{\textrm{1min}} < E_1^* < E_{\textrm{1max}}\,.$ This energy distribution can be divided into infinitely many delta functions all next to each other. Boosting each one of them separately creates a rectangular distribution. A delta function centered on energy $E_1^*$ will become a rectangle of width $2\gamma\beta\sqrt{(E_1^*)^2 - m_f^2}$ centered at $\gamma E_1^*$. This produces a series of overlapping rectangles stacked next to each other. To get the total spectrum, we have to sum up the contributions of each rectangle. 

\begin{figure}[tbp]
\centering
\begin{minipage}{0.48\textwidth}
  \centering
  \includegraphics[width=\linewidth]{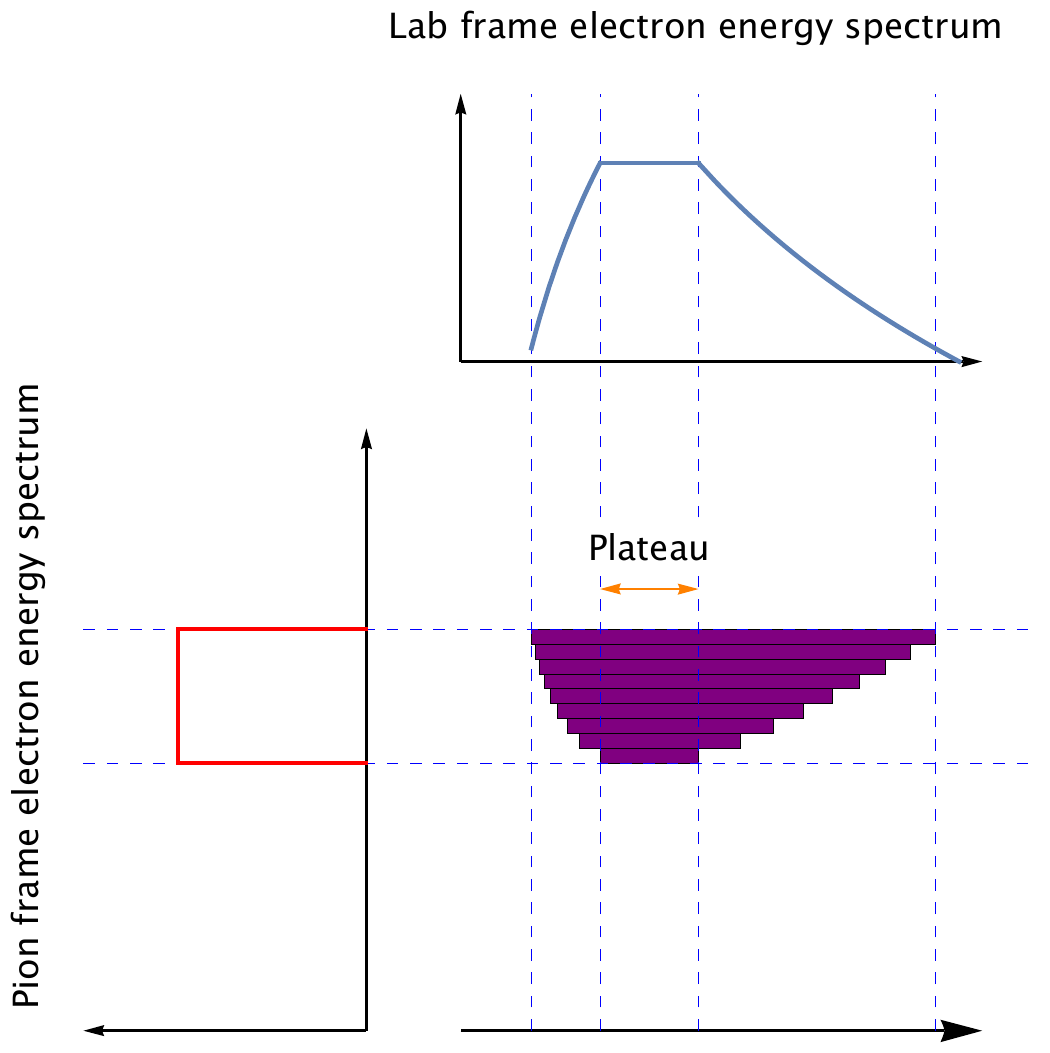}
  \label{fig:plateau_high}
\end{minipage}
\hfill
\begin{minipage}{0.48\textwidth}
  \centering
  \includegraphics[width=\linewidth]{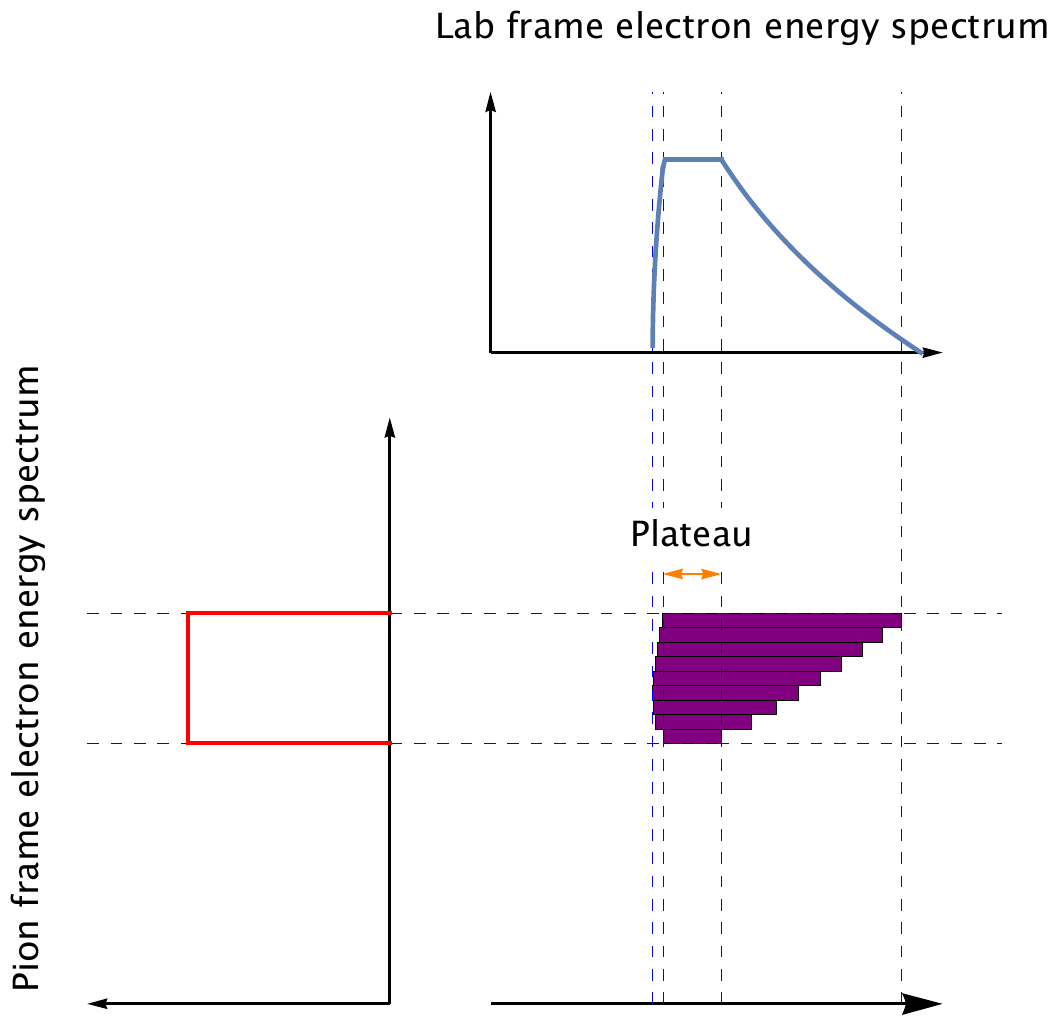}
  \label{fig:plateau_low}
\end{minipage}
\caption{Illustration of how plateaus emerge from boosted spectra in the case of a large boost (left) and a small boost (right). The pion-frame electron energy distribution (red) is boosted into the lab-frame distribution (blue). This may be understood by cutting up the pion frame spectrum into nearly mono-energetic slices, each of which is then boosted into a purple rectangle in the center. Vertically summing up the purple rectangles yields the lab frame spectrum on top, which has a plateau at energies where all of the purple rectangles overlap. The four important points of the spectrum, \cref{eq:spec1,eq:spec2,eq:spec3,eq:spec4}, are denoted by vertical dashed blue lines.}\label{fig:plateau_explanation}
\end{figure}

This reasoning is illustrated in \cref{fig:plateau_explanation}. The red graph is the pion frame electron energy spectrum without spin correlations. The purple rectangles represent what happens when the pion frame spectrum is chopped up and boosted. We now see that the plateau originates from the mutual overlap of purple rectangles. This analysis yields four important points of the boosted spectrum. The spectrum usually begins at 
\begin{align}
E_1 = \gamma\left(E_{1\textrm{min}} - \beta\sqrt{E_{1\textrm{min}}^2 - m_f^2} \ \right)\,,\label{eq:spec1}
\end{align}
unless $\gamma m_f > E_{\textrm{1min}}$ then, it starts at the fermion mass $m_f$. The spectrum gradually grows until it reaches a plateau which begins at 
\begin{align}
E_1 = \textrm{Min}\left(\gamma\left(E_{1\textrm{min}} + \beta\sqrt{E_{1\textrm{min}}^2 - m_f^2}\right)\,, \gamma\left(E_{1\textrm{max}} - \beta\sqrt{E_{1\textrm{max}}^2 - m_f^2}\right)\right)\label{eq:spec2}
\end{align}
and ends at 
\begin{align}
E_1 = \textrm{Max}\left(\gamma\left(E_{1\textrm{min}} + \beta\sqrt{E_{1\textrm{min}}^2 - m_f^2}\right)\,, \gamma\left(E_{1\textrm{max}} - \beta\sqrt{E_{1\textrm{max}}^2 - m_f^2}\right)\right)\,.\label{eq:spec3}
\end{align}
Finally, the whole spectrum must end at 
\begin{align}
E_1 = \gamma\left(E_{1\textrm{max}} + \beta\sqrt{E_{1\textrm{max}}^2 - m_f^2} \ \right)\,.\label{eq:spec4}
\end{align}

\bibliographystyle{utphys}
\bibliography{spincorrelations}

\end{document}